\documentclass{aa}  

\usepackage{graphicx}
\usepackage{txfonts}
\usepackage{pdflscape}
\usepackage{xcolor}
\usepackage{booktabs}
\usepackage{hyperref}
\usepackage{multirow}
\renewcommand{\arraystretch}{1.2}
\hypersetup{
  colorlinks=true,
  citecolor=blue,
  urlcolor=blue,
  linkcolor=blue
}
\usepackage[separate-uncertainty = true,multi-part-units=single,group-separator={}, group-digits = integer, group-four-digits = true,per-mode=reciprocal]{siunitx}
\DeclareSIUnit\parsec{pc}
\usepackage{chemformula}
\usepackage{threeparttable} 

\newcommand{\kms}{\mbox{${\rm km\,s}^{-1}$}\xspace}

\newcommand{\Mjup}{\mbox{${\rm M}_\mathrm{J}$}}

\newcommand{\Searth}{\mbox{${S}_{\oplus}$}}

\newcommand{\Rjup}{\mbox{${\rm R}_\mathrm{J}$}}

\newcommand{\vsini}{\mbox{$v \sin i_{\ast}$}\xspace}

\newcommand{\halpha}{\mbox{H$\alpha$}\xspace}
\newcommand{\hbeta}{\mbox{H$\beta$}\xspace}
\newcommand{\hgamma}{\mbox{H$\gamma$}\xspace}

\newcommand{\NaI}{\ion{Na}{i}\xspace}
\newcommand{\MgI}{\ion{Mg}{i}\xspace}
\newcommand{\MgII}{\ion{Mg}{ii}\xspace}
\newcommand{\LiI}{\ion{Li}{i}\xspace}
\newcommand{\FeI}{\ion{Fe}{i}\xspace}
\newcommand{\FeII}{\ion{Fe}{ii}\xspace}
\newcommand{\water}{\mbox{H$_2$O}\xspace}
\newcommand{\Htwo}{\mbox{H$_2$}\xspace}
\newcommand{\target}{WASP-178~b\xspace}

\newcommand{\numpm}[3]{$#1^{+#2}_{#3}$}

\newcommand{\CONAN}{{\sc \tt CONAN}\xspace}

\begin{document}

   \title{The atmospheric composition of the ultra-hot Jupiter \target observed with ESPRESSO\thanks{Based on Guaranteed Time Observations collected at the European Southern Observatory under ESO programme 1104.C-0350 by the ESPRESSO Consortium.}}

    \authorrunning{Damasceno et al. (2024)}

   \author{Y.\,C.\,Damasceno\inst{1,}\inst{2,}\inst{3}, J.\,V.\,Seidel\inst{3}\thanks{ESO Fellow}, B.\,Prinoth\inst{3,}\inst{4}, A.\,Psaridi\inst{5}, E.\,Esparza-Borges\inst{6,}\inst{7}, M.\,Stangret\inst{8}, N.\,C.\,Santos\inst{1,}\inst{2}, M.\,R.\,Zapatero-Osorio\inst{9}, Y.\,Alibert\inst{10}, R.\,Allart\inst{11,}\inst{5}\thanks{Trottier Postdoctoral Fellow}, T.\,Azevedo Silva\inst{1,}\inst{2}, M.\,Cointepas\inst{5,}\inst{12}, A.\,R.\,Costa Silva\inst{1,}\inst{2,}\inst{5}, E.\,Cristo\inst{1,}\inst{2}, P.\,Di\,Marcantonio\inst{13}, D.\,Ehrenreich\inst{5,}\inst{14}, J.\,I.\,Gonz\'alez Hern\'andez\inst{6,}\inst{7}, E.\,Herrero-Cisneros\inst{15}, M.\,Lendl\inst{5}, J.\,Lillo-Box\inst{16}, C.\,J.\,A.\,P.\,Martins\inst{1,17}, G.\,Micela\inst{18}, E.\,Pall\'e\inst{6,}\inst{7}, S.\,G.\,Sousa\inst{1}, M.\,Steiner\inst{5}, V.\,Vaulato\inst{5}, Y.\,Zhao\inst{5}, F.\,Pepe\inst{5}}

   \institute{Instituto de Astrofísica e Ci\^encias do Espa\c{c}o, Universidade do Porto, CAUP, Rua das Estrelas, 4150-762 Porto, Portugal 
   \and
   Departamento de F\'isica e Astronomia, Faculdade de Ci\^encias, Universidade do Porto, Rua
do Campo Alegre, 4169-007 Porto, Portugal 
   \and
   European Southern Observatory, Alonso de C\'ordova 3107, Vitacura, Regi\'on Metropolitana, Chile
   \and Lund Observatory, Division of Astrophysics, Department of Physics, Lund University, Box 118, 221 00 Lund, Sweden 
   \and Observatoire de Gen\`eve, D\'epartement d’Astronomie, Universit\'e de Gen\`eve, Chemin Pegasi 51, 1290 Versoix, Switzerland  
    \and Instituto de Astrof\'isica de Canarias, E-38200 La Laguna, Tenerife, Spain 
    \and Departamento de Astrof\'isica, Universidad de La Laguna, E-38206 La Laguna, Tenerife, Spain 
    \and INAF – Osservatorio Astronomico di Padova, Vicolo dell'Osservatorio 5, 35122, Padova, Italy 
   \and Centro de Astrobiolog\'{i}a, CSIC-INTA, Camino Bajo del Castillo s/n, 28692 Villanueva de la Ca\~nada, Madrid, Spain 
   \and Center for Space and Habitability, University of Bern, Gesellsschaftsstr. 6, 3012 Bern, Switzerland 
   \and D\'epartement de Physique, Institut Trottier de Recherche sur les Exoplan\`etes, Universit\'e de Montr\'eal, Montr\'eal, Qu\'ebec, H3T 1J4, Canada 
   \and Univ. Grenoble Alpes, CNRS, IPAG, F-38000 Grenoble, France 
   \and INAF – Osservatorio Astronomico di Trieste, via G. B. Tiepolo 11, I-34143, Trieste, Italy 
   \and Centre Vie dans l’Univers, Facult\'e des sciences de l’Universit\'e de Gen\`eve, Quai Ernest-Ansermet 30, 1205 Geneva, Switzerland 
   \and Centro de Astrobiolog\'ia, CSIC-INTA, Crta. Ajalvir km 4, E-28850 Torrej\'on de Ardoz, Madrid, Spain 
   \and Centro de Astrobiolog\'ia (CAB, CSIC-INTA), ESAC campus, 28692, Villanueva de la Ca\~nada (Madrid), Spain 
   \and Centro de Astrof\'isica da Universidade do Porto, Rua das Estrelas, 4150-762 Porto, Portugal 
   \and INAF - Osservatorio Astronomico di Palermo, Piazza del Parlamento 1, 90134 Palermo, Italy 
}

   \date{Received ...}

 
  \abstract
   {Ultra-hot Jupiters (UHJ) have emerged as ideal testbeds for new techniques for studying exoplanet atmospheres. Only a limited number of them are currently well studied, however.}
   {We search for atmospheric constituents for the UHJ \target with two ESPRESSO transits. Additionally, we show parallel photometry that we used to obtain updated and precise stellar, planetary, and orbital parameters.}
   {The two transits we obtained were analysed with narrow-band transmission spectroscopy and with the cross-correlation technique to provide detections at different altitude levels. {We focused on searching for \NaI, \halpha, \hbeta, \hgamma, \MgI, and \LiI lines in narrow-band data}, as well as \FeI and \FeII, and attempted to confirm \MgI with the cross-correlation technique. We corrected for the Rossiter-McLaughlin effect and regions with a low signal-to-noise ratio due to \NaI absorption in the interstellar medium. We then verified our results via bootstrapping.}
   {We report the resolved line detections of \NaI ($5.5\,\sigma$ and $5.4\,\sigma$), \halpha ($13\,\sigma$), {\hbeta ($7.1~\sigma$)}, and tentatively \MgI ($4.6\,\sigma$). With a cross-correlation, we confirm the \MgI detection ($7.8\,\sigma$ and $5.8\,\sigma$), and we additionally report the detections of \FeI ($12\,\sigma$ and $10\,\sigma$) and \FeII ($11\,\sigma$ and $8.4\,\sigma$) on both nights separately. The detection of \MgI remains tentative, however, because the results on the two nights differ. The results also differ compared with the properties derived from the narrow-band data.
   }
   {None of our resolved spectral lines probing the middle to upper atmosphere shows significant shifts relative to the planetary rest frame. \halpha{} {and \hbeta exhibit a respective line broadening of $39.6\pm 2.1~\kms$ and $27.6\pm 4.6~\kms$, however,} indicating the onset of possible escape. \target differs from similar UHJ by its lack of strong atmospheric dynamics in the upper atmosphere. The broadening seen for \FeI (\num{15.66(0.58)} \kms) and \FeII (\num{11.32(0.52)} \kms) might indicate the presence of {winds in the mid-atmosphere, however}. {Future studies of the impact of the flux variability caused by the host star activity might shed more light on the subject}. {Previous work indicated the presence of SiO cloud-precursors in the atmosphere of WASP-178 b and a lack of \MgI and \FeII. However, our results suggest that a scenario in which the planetary atmosphere is dominated by \MgI and \FeII  is more likely.} In light of our results, we encourage future observations to further elucidate these atmospheric properties.
   }


    \maketitle
%

\section{Introduction}

\begin{figure}
        \centering
        \includegraphics[width=\linewidth]{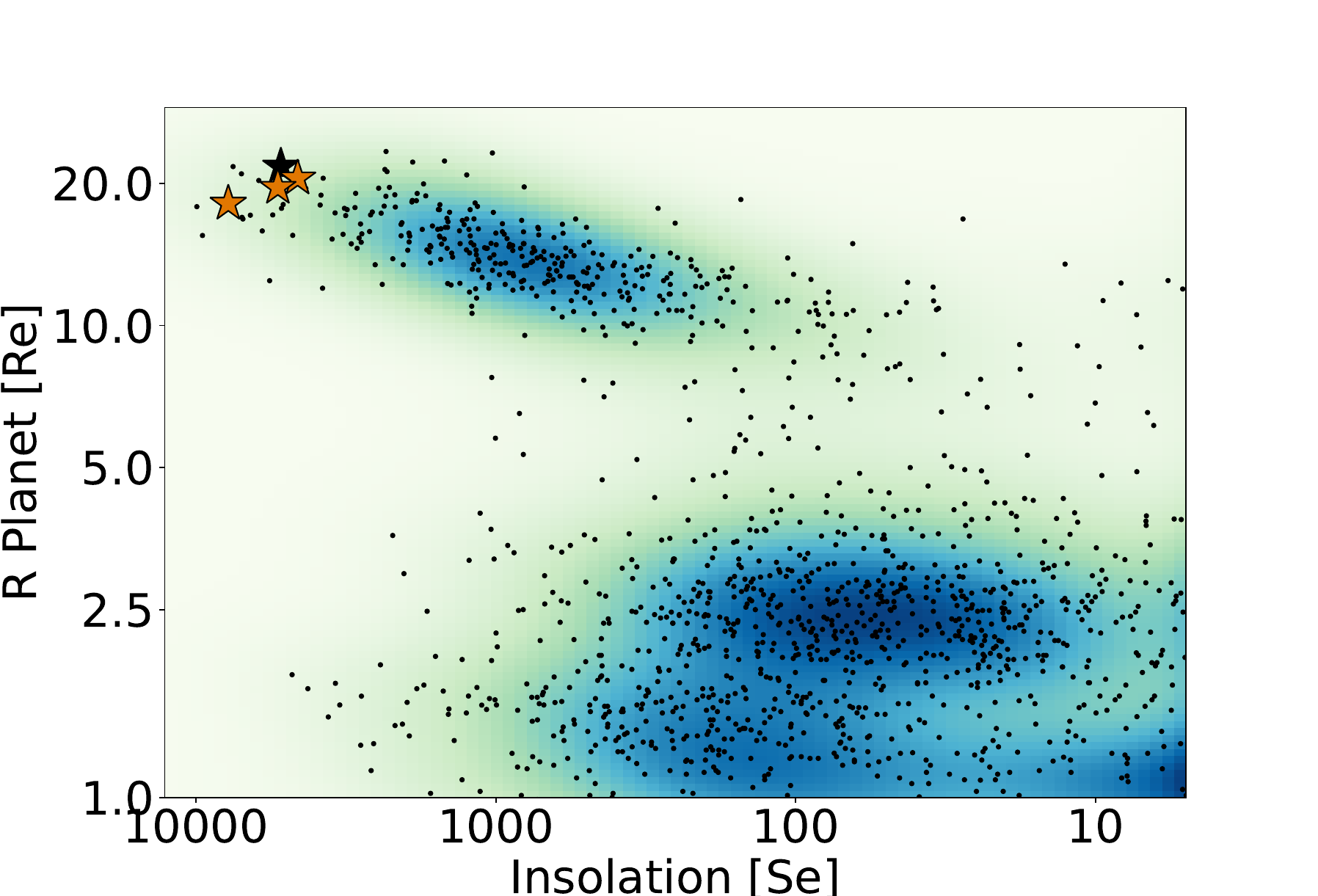}
        \caption{Exoplanets in the radius vs. insolation plot, in terms of Earth's values. The {UHJ} \target is marked with a black star. WASP-121 b, WASP-76 b, and WASP-189 b, mentioned in the introduction, are marked as orange stars.}
        \label{fig:Re_inso}
    \end{figure}

\noindent Ultra-hot Jupiters (UHJ) are defined as Jupiter-like planets with equilibrium temperatures above 2000~K. They have emerged as ideal testbeds within the exoplanet community. Due to their close-in orbit around their host star, their atmospheres tend to be dissociated into atoms or are even ionised, proportioning a bloated atmosphere where the depth and width of the absorptions can be translated into height and velocity distribution. The inspection of multiple elements that sit in different regions of the atmosphere allows for the holistic exploration of the chemistry, temperature-pressure profile, and atmospheric dynamics of these planets directly from data. A range of modelling and retrieval techniques were either benchmarked or matured using data from UHJ, for instance, the detection of a wide variety of chemical species \citep{Hoeijmakers2019, Prinoth2022}, the retrieval of abundances \citep{Gandhi2022, Gibson2022,Pelletier2023}, the observation of emission features \citep{Pino2020, Pino2022}, and the observation of atmospheric dynamics \citep{Borsa2019, Ehrenreich2020, Seidel2021, Seidel2023}.
However, due to their extreme conditions, only a handful of suitable targets for atmospheric characterisation fall into the category of UHJs. Among the dozens of UHJ whose atmospheres have already been observed, the most canonically studied are WASP-121 b \citep{Delrez2016, AzevedoSilva2022, Seidel2023}, WASP-76 b \citep{West2016, Tabernero2021}, WASP-189 b \citep{Lendl2020, Stangret2022, Prinoth2022}, and the hottest exoplanet to date, KELT-9 b \citep{Gaudi2017, Hoeijmakers2018, Wyttenbach2020,Borsato2023}.
\target has an equilibrium temperature of $2470\pm60$ K, with a planetary radius of $1.81\pm0.09~\Rjup$, a mass of $1.66\pm0.12~\Mjup$, and an insolation of \numpm{5550}{1250}{-630} $\Searth$ \citep{Hellier2019, Rodriguez2020} (see Fig. \ref{fig:Re_inso}, \target is marked as a black star). For context, this places \target between WASP-121~b and WASP-76~b in terms of insolation. However, differing from these two targets, WASP-178 is the second hottest star known to date to host an exoplanet, a type A1 IV-V star with an effective stellar temperature of $9350\pm150$ K \citep{Hellier2019} and with an apparent magnitude in the visible band of $9.95$. \\

\target has been studied previously with {the CHaracterising ExOPlanets Satellite (CHEOPS) and the Transiting Exoplanet Survey Satellite (TESS)} and was found to be in a near pole-on geometry, and the rotation period of the star is similar to the orbital period of the planet \citep{Pagano2023}. {\citet{Pagano2023} found} efficient heat transport from the day to the night side, thus observationally challenging the current theory that heat transport by zonal winds becomes less efficient with increasing equilibrium temperature \citep[e.g][]{PerezBecker2013, Komacek2017, Schwartz2017, Parmentier2018}. Additionally, \target was observed with HST/WFC3/UVIS\footnote{Hubble Space Telescope/Wide Field Camera 3/Ultraviolet Imaging Spectrograph}, and one of the largest {near ultraviolet (NUV)} spectral features to date \citep{Lothringer2022}. {No transit asymmetry was found by \citet{Lothringer2022}}, indicating that signals are present on both terminators. Photochemical hazes are ruled out due to a lack of scattering slope, and {the} observed spectral features can be explained by either SiO at solar abundance or by an atmosphere that is dominated by \MgI and \FeII at super-solar abundances without SiO. Surprisingly, \citet{Lothringer2022} report non-detections for \FeI and \FeII.

In this work, we present two transit observations of \target obtained with the {Echelle SPectrograph for Rocky Exoplanets and Stable Spectroscopic Observations spectrograph (ESPRESSO)}. The manuscript is structured as follows: In Sect. \ref{sec:obs} we report the data quality of the two transits and our data reduction, and Sect. \ref{sec:phot} analyses the parallel photometry taken with the EulerCam at the Swiss 1.2m Euler telescope. Sections \ref{sec:transSpec} and \ref{sec:crosscorr} show the narrow-band and cross-correlation transmission analysis and detections, respectively, which are then placed into context with the literature in Sect. \ref{sec:diss}.


\section{ESPRESSO observations and data analysis}
\label{sec:obs}
    We observed the WASP-178 system and captured two full transits of the {UHJ} \target. The target was observed with ESPRESSO \citep{Pepe2021} in HR21 mode ($\mathcal{R}~\approx 140.000$), which is installed at the ESO VLT telescopes at the Paranal Observatory in Chile. The transits were observed on the nights of 2021 May 3 and 2021 July 9 with the UT4 and UT2 telescopes, respectively. Both observations were performed with fibre A on the target and fibre B on the sky to monitor telluric emission. 

\begin{figure}
        \centering
        \includegraphics[width=\linewidth]{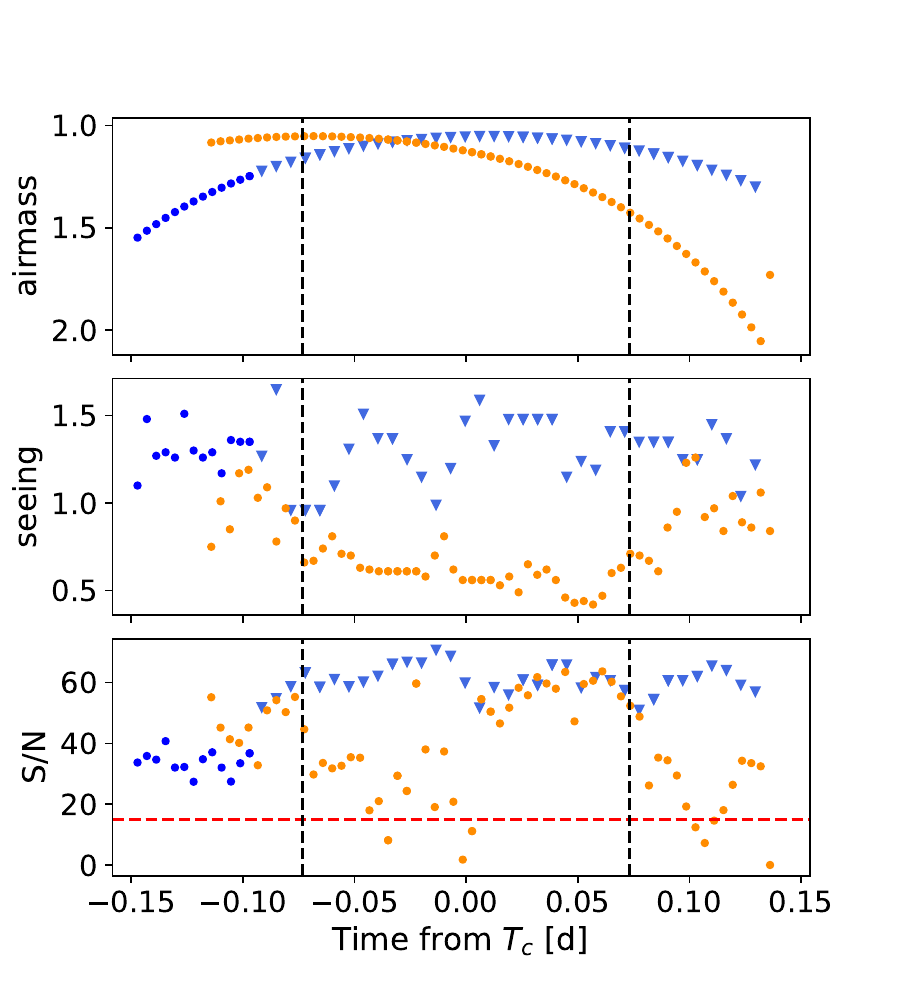}
        
        \caption{Conditions of the first (blue markers) and second (orange markers) observed transits {showing the airmass (top panel), seeing (middle panel), and the S/N (bottom panel) for each exposure}. The S/N was calculated from the \NaI D doublet orders. The dashed black lines show the beginning and end of transit. The dashed red line marks the S/N = $15$ threshold, below which exposures were discarded. The circular markers correspond to $300$\,s exposures, and the lighter shaded triangular markers are used for $500$\,s exposures.}
        \label{fig:conditions_1}
    \end{figure}

    On the first night, $48$ spectra were taken. The first $13$ spectra had an exposure time of $300$~s. Because of a consistently high seeing, the remaining $35$ spectra had exposure times of $500$~s. On the second night, $61$ spectra were taken, all with an exposure time of $300$~s. In total, the two nights resulted in $109$ spectra, $58$ of which were taken in-transit, and $51$ spectra were out of transit. In both nights, the transits were observed from start to finish. The first night presented a higher signal-to-noise ratio (S/N) consistency. The airmass, seeing, and S/N of each exposure are shown in Fig. \ref{fig:conditions_1} and in Table \ref{tab:observations}. The S/N shown in both {is that of the spectral order $116$}. The spectra were reduced using the ESPRESSO data reduction pipeline (DRS version 3.0.0), queried from the DACE\footnote{https://dace.unige.ch/} platform, and the orders were extracted and corrected for the blaze function. We worked in the narrow band with the S2D blaze-corrected and sky-subtracted spectra. We focused on the paired orders containing the \NaI doublet ($116$ and $117$), the \halpha line ($138$ and $139$), {the \hbeta line ($70$ and $71$), the \hgamma line ($40$ and $41$)}, the \MgI b triplet ($86$ and $87$), and the \LiI resonant line ($142$ and $143$). We also searched throughout the entire wavelength range of ESPRESSO for \FeI, \FeII, and \MgI{} in cross-correlation using the S2D spectra without blaze correction or sky subtraction.

   \begin{table*}
        \caption{Overview of the two transit observations for \target}\vspace{-1em}
        \begin{center}
                \begin{tabular}{llllll}
                        \toprule
                        Date   & $\#$ Spectra$^1$  & Exp. Time [s] & Airmass & Seeing & S/N $\#$116 \\
                        \midrule
                        2021-03-05   & 48 (23/25)  & 300 and 500  & 1.55 - 1.05 - 1.30 & 1 to 1.7 & 27 to 71 \\
                        2021-07-09   & 61 (35/26)$^2$  & 300  & 1.08 - 1.05 - 2.05 & 0.4 to 1.2 & 0 to 64 $^3$ \\
                        \bottomrule
                \end{tabular}
        \end{center}
        \vspace{-1em}
        \textit{Note:} $^{1}$ The in- and out-of-transit spectra are listed in parentheses, respectively.  $^2$ Seven pectra were discarded due to low S/N, leaving $32$ in-transit and $22$ out-of-transit spectra.  $^3$ After the low S/N spectra were removed, the lowest S/N value was $18$.
        \label{tab:observations}
\end{table*}

\subsection{Data reduction}
    The spectra from both transits, as reduced via the ESPRESSO DRS, are given in the reference frame of the Solar System barycentre, and the wavelength is given in air and corrected for the blaze.

    The second night of observations was affected by clouds, resulting in seven exposures with low S/N ($\leq$ $15$). As a consequence, we discarded these spectra. This left a total of $102$ spectra, $55$ in-transit and the remaining $47$ out of transit. The data reduction done by the ESPRESSO DRS corrects for most cosmic rays, but a few are still present in the reduced data. In each exposure, we rejected the remaining cosmic rays using a $6$ $\sigma$ clipping routine on the flux difference between adjacent wavelength bins. The rejected points were then replaced by the mean of the six nearest values. Between all exposures from both nights, we replaced a total of {$24$ data points in orders $50$ and {$51$}, $54$ points in orders $70$ and $71$,} $24$ points in orders $86$ and $87$, $75$ points in orders $116$ and $117$, $65$ points in orders $138$ and $139$, and $75$ points in orders $142$ and $143$. This means that overall, less than $0.0002~\%$ of the reduced data in these orders were affected.

    \subsection{Telluric correction}

    Ground-based observations are subject to contamination from Earth’s atmosphere. Telluric contamination is time variable and depends on the airmass and water-vapour column in the line of sight. We performed the telluric correction using \texttt{molecfit} version 1.5.1 \citep{Smette2015, Kausch2015}, an ESO software developed for modelling the Earth's spectral features for ground-based observations. This approach to telluric absorption correction has been used in the exoplanetary field in high-resolution spectroscopy data, such as spectra from the High Accuracy Radial velocity Planet Searcher (HARPS) \citep{Allart2017} and from ESPRESSO \citep{Allart2020,Tabernero2021,Seidel2022}. For further details on the \texttt{molecfit} application to exoplanet spectra, we refer to \citet{Allart2017}.
    
\begin{figure*}
        \sidecaption
        \includegraphics[width=12cm]{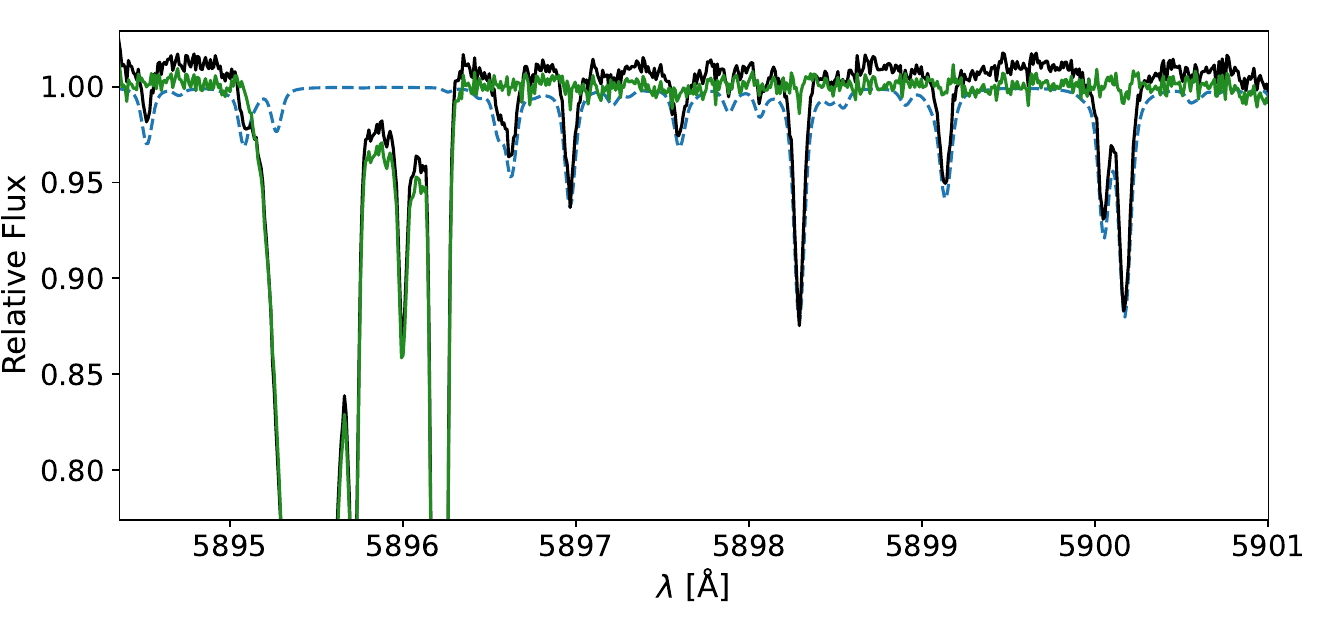}
        \caption{Mean spectrum before ({solid} black line) and after (solid green line) the telluric correction, for the first night of observations in the Earth's rest frame. The telluric model averaged over all exposures is represented by the dashed blue line.}
        \label{fig:tellurics}
    \end{figure*}

    In the spectral range probed by ESPRESSO, the elements that cause most of the contamination are \water{}, O$_2$, and O$_3$. Considering the orders addressed in this work, we only included the contributions from \water{} when we computed the telluric model spectra because {O$_3$ affects the visible wavelength range via broadband absorption} and the O$_2$ lines were masked for the cross-correlation analysis. The telluric model for \water was fitted on the wavelength range near the sodium doublet, in orders $115$ and $116$. In these orders, the telluric lines are weak, and the fitting region therefore had to be carefully selected such that it did not include any stellar features. The fitted telluric model was then applied to all orders. For the spectral regions studied in the narrow band, this correction was sufficient, while for the cross-correlation analysis, we corrected for O$_2$ by masking its respective features. Figure \ref{fig:tellurics} shows the mean spectrum over all exposures, normalised for better visibility, before and after the telluric correction, as well as the model telluric spectrum. Fig. \ref{fig:tellurics} shows that the telluric lines are corrected down to the noise level. We additionally verified from fibre B that no emission from telluric sodium was present during our observations.
    
    \subsection{\NaI{} absorption in the interstellar medium} 

\begin{figure}
    \centering
    \includegraphics[width=\linewidth]{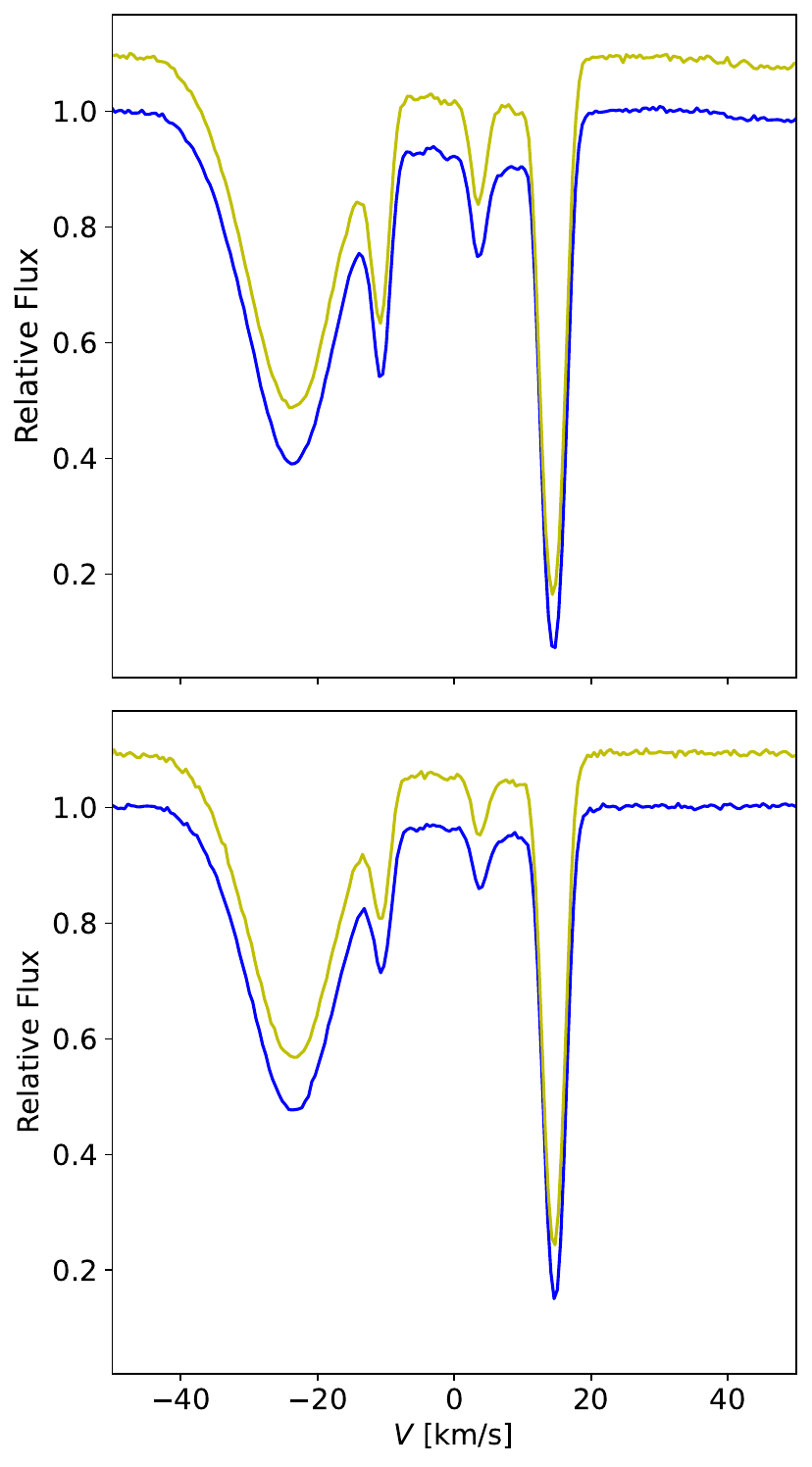}
        \caption{Average normalised spectrum for the first ({solid} blue line) and second ({solid} yellow line) transits, showing the vicinity of the sodium D2 (top {panel}) and D1 (bottom {panel}) lines. The yellow line is additionally offset for better visibility. The spectrum is shown as a function of the Doppler velocity with reference to the rest-frame wavelength of each line in the Solar System barycentre.}
        \label{fig:ISM}
    \end{figure}

    Near the stellar \NaI Fraunhofer lines, we identified {three additional absorption features}. The stellar lines are shifted {by \num{-23.43(0.02)} \kms} with respect to the Solar System barycentre rest frame, which is consistent with the expected system velocity reported in \citet{Hellier2019}, considering the $0.5$ \kms bin size of ESPRESSO. During the first transit, the three rightmost lines are shifted by {\num{-10.41(0.02)}, \num{4.10(0.09)}, and \num{14.72(0.01)} \kms, and on the second transit, the velocities are \num{-10.42(0.03)}, \num{3.88(0.12)}, and \num{14.72(0.01)} \kms}. The line centres were estimated through a least-squares Gaussian fit. Each pair of lines was shifted equally from the expected position of the doublet on both nights, indicating they are not of telluric origin. In Fig. \ref{fig:ISM} the spectrum portions near the sodium doublet are shown in velocity space with respect to each of the doublet lines. Because these lines follow the stellar features, they are likely caused by separate interstellar medium (ISM) clouds with different radial velocities. 

    We treated the ISM lines the same way as the stellar lines by including them in the master out-of-transit spectrum. Due to the very low S/N around the deepest ISM line, we masked a $0.22$ $\AA$ band around its line centre. The masked values were replaced with the local continuum, which was determined as the mean of the $1$ $\AA$ band towards the red of the mask since the adjacent points towards the blue contain another ISM absorption line. Based on the calculated planet and star velocities, no planetary sodium absorption feature is expected to cross the masked region in any of the exposures. This process is therefore not expected to affect the results.

    \subsection{Modelling the Rossiter-McLaughlin effect}
The Rossiter-McLaughlin (RM) effect \citep{Holt1893,Rossiter1924,McLaughlin1924} is a result of the non-uniformity throughout the stellar surface, with contributions from the limb darkening and stellar rotation, which affects the in-transit data as the planet passes through different portions of the stellar disk \citep{CasasayasBarris2020}. This effect can be measured from the radial velocities (RVs) as they deviate from the Keplerian motion of the star during a transit. The exact pattern imprinted on the RVs and the spectra depends on the orbit architecture as well as on the stellar and planetary parameters \citep{Triaud2018}. 

We studied the obliquity of the system by analysing the RM signal imprinted on the RV measurements and measuring the sky-projected spin-orbit angle $\rm \lambda$. We combined the RV measurements of both transits using a $\rm 14.4~min$ binning and extracted the RM signal by subtracting a linear fit from the out-of-transit RVs. The approach used by the ESPRESSO DRS pipeline to measure the RV is based on the fitting of a Gaussian to the cross-correlation function (CCF) \citep[find this technique detailed in ][]{Baranne1996, Pepe2002}. Accordingly, we analysed the RM signal using a Python implementation of \texttt{ARoME}\footnote{\texttt{ARoME} code is publicly available at \url{http://www.astro.up.pt/ resources/arome/}} \citep{Boue2013}. This code was specially designed to model the RV data extracted by the CCF approach. Based on \texttt{ARoME} models and the parameters from \citet{Hellier2019}, we performed a Markov chain Monte Carlo (MCMC) fitting of the RM signal with \texttt{emcee} \citep{emcee} using 11 chains of 5000 steps each. During the fitting, we left the mid-transit time ($\rm T_{0}$), the sky-projected spin-orbit angle ($\rm \lambda$), and the projected stellar rotational velocity ($v\sin{i}$) as free parameters. We imposed wide uniform priors on the three free parameters: $\rm -0.02~days<T_{0}<0.02~days$, $\rm -90~deg<\lambda<160~deg$, and $\rm -60~kms^{-1}<v\sin{i}<60~kms^{-1}$. Additionally, we fixed the remaining planetary, stellar, and instrumental parameters in the fitting, including the limb-darkening coefficients that were estimated through \texttt{LDTk} \citep{LDTk}.

\begin{figure}
    \centering
    \includegraphics[width=\linewidth]{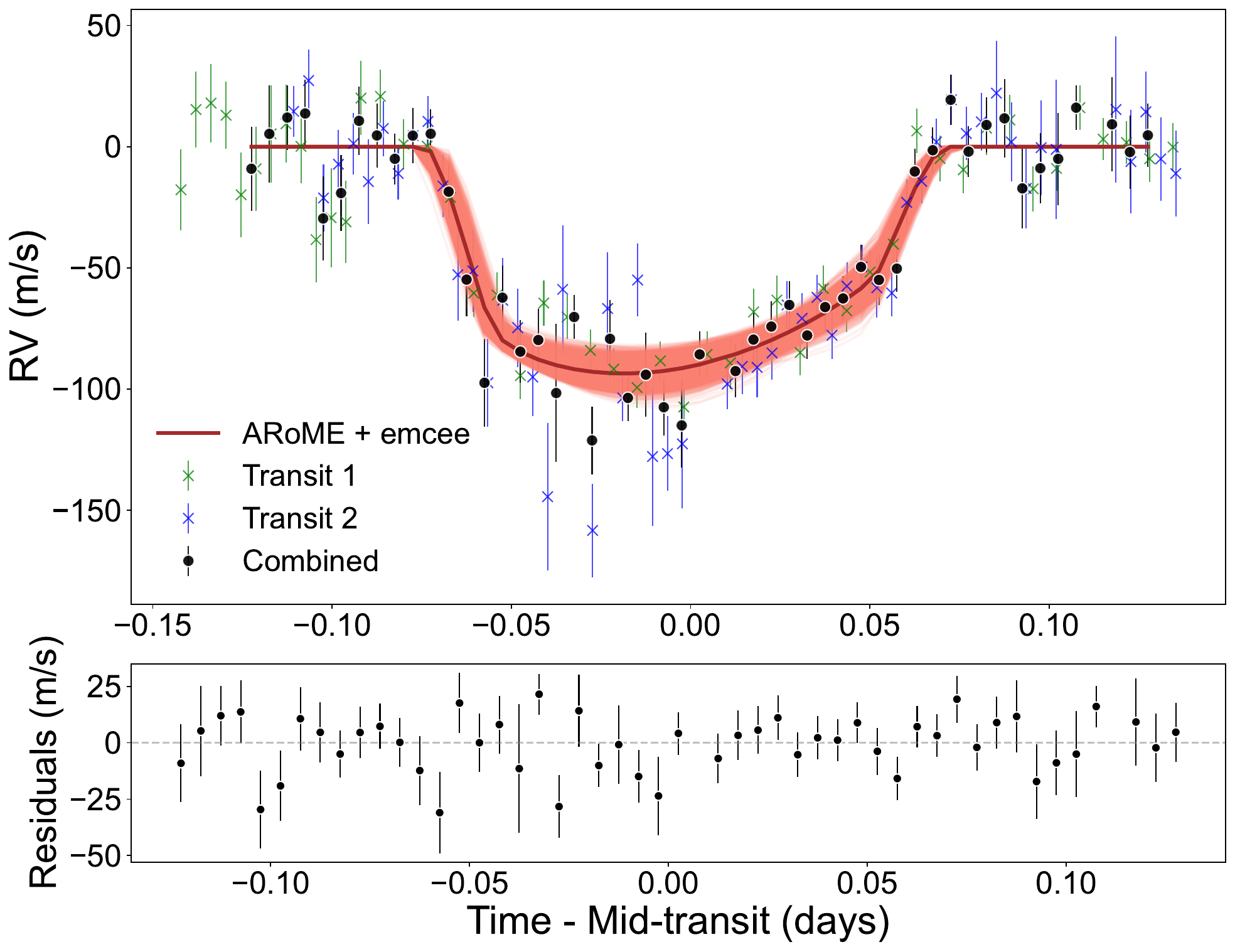}
        \caption{Rossiter-McLaughlin signal imprinted on the first transit (green crosses), second transit (blue crosses), and combined (black dots) RV time series. Additionally, we show {in the top panel} the best RM model from the \texttt{ARoME} + \texttt{emcee} fitting (magenta) and models within $\rm 1\sigma$ (light pink). {The bottom panel contains the residuals of the combined RV time series.}}
        \label{fig:RM_fitting_Arome}
    \end{figure}

\begin{figure}
    \centering
    \includegraphics[width=\linewidth]{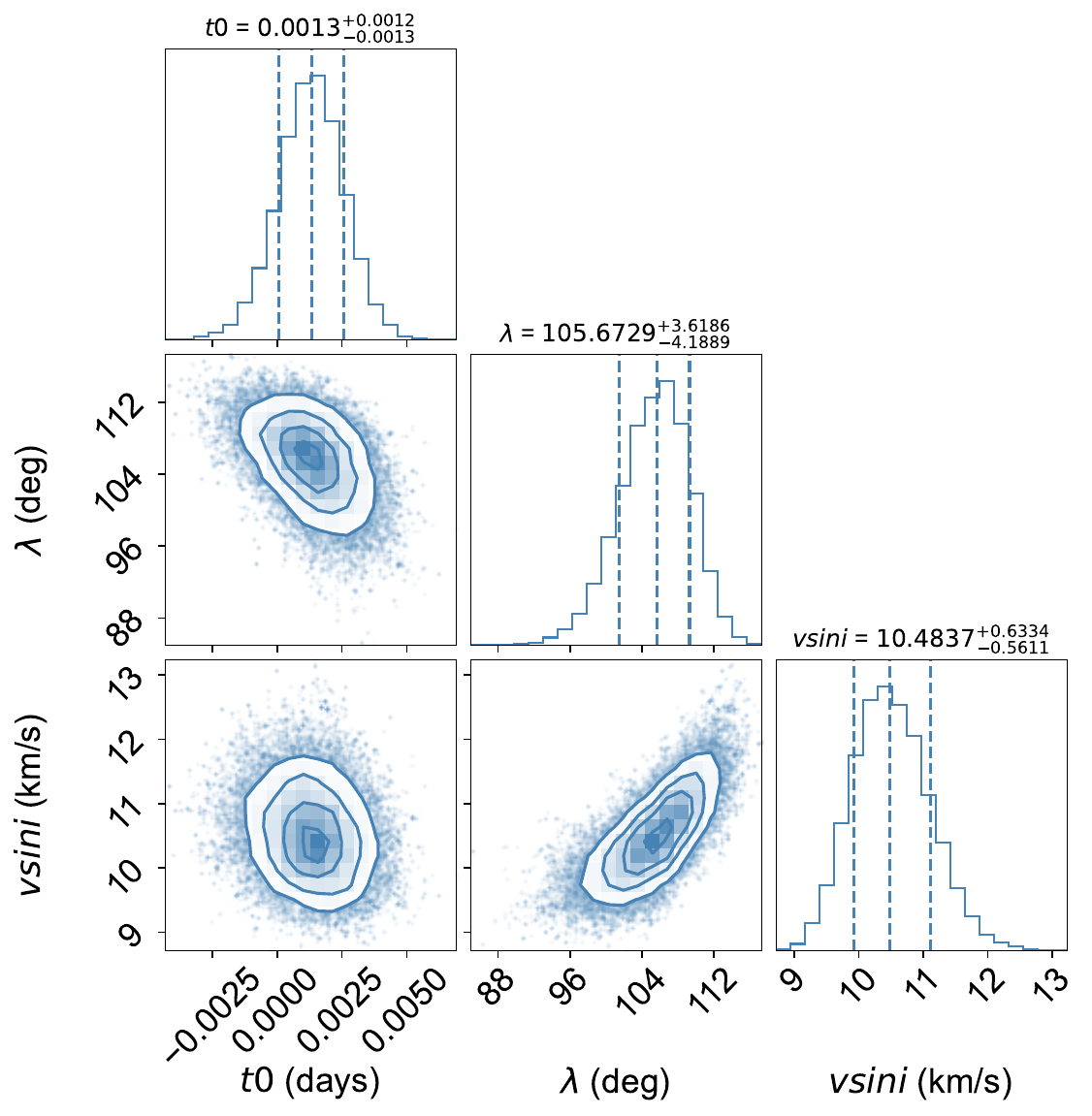}
        \caption{Resulting posterior distributions for the three free ($\rm T_{0}$, $\rm \lambda$, and $v\sin{i}$) parameters from the RM fitting.}
        \label{fig:Corner_Arome}
    \end{figure}

As a result of the RM analysis, we found \target to be a misaligned planet, measuring $\rm \lambda = 105.7 ^{+3.6}_{-4.2}~deg$, which is consistent with the value presented in \citet{Rodriguez2020} within the $3~\sigma$ level. We show the best-fit RM model and RM models within $\rm 1~\sigma$ in Fig.~\ref{fig:RM_fitting_Arome}. We present the posterior distributions from the fitting in Fig.~\ref{fig:Corner_Arome}. {The stellar \vsini result shows a $3~\sigma$ difference from the literature value of \num{8.2(0.6)} \kms \citep{Hellier2019}, which can be associated with the use of updated system parameters. The effect of this difference in the RM profiles is small and therefore has no notable impact on our results}. The unusually low rotational velocity for a spectral type A star might be explained if we are observing it on a near pole-on position. This hypothesis was posed in \citet{Rodriguez2020} and was supported by results in \citet{Pagano2023}.

We corrected for the RM effect using the code \texttt{StarRotator}\footnote{https://github.com/Hoeijmakers/StarRotator}, which estimates the star's spectrum during a transit via numerical integration of the stellar disk, allowing us to obtain a model for the stellar spectrum behind the planet at each phase. We generated the simulated star in a $400\times400$ grid without differential rotation and considered the limb-darkening coefficients determined from the simultaneous photometry. The model spectra were taken from the PHOENIX database \citep{Husser2013} for each observed phase of the transit and included the centre-to-limb variations with a quadratic limb-darkening law. The RM effect at each input phase was obtained by dividing the in-transit spectra by the out-of-transit spectra and convolving with the instrumental resolution (considered $140.000$ for the ESPRESSO mode). We used the stellar and planetary parameters listed in Table \ref{tab:parameters_W178}, except for T$_\mathrm{eff}$, log(g), and [Fe/H], which we rounded up to the nearest allowed values, $9400$ K, 4.5 dex (in cgs units) and 0.0 dex (in solar metallicity), respectively. The resulting residuals were then used to correct for the RM effect on all in-transit exposures.


\section{Simultaneous photometry with EulerCam} 
\label{sec:phot}
We observed four full transits of \target with the EulerCam instrument \citep{Lendl2012} installed on the 1.2~m Euler telescope in La Silla. The transits were observed in the $r’-Gunn$ filter on 2021 May 3, 2021 June 29, 2021 July 9, and 2023 April 2 with exposure times of 110~s, 40~s, 42~s, and 33~s, respectively. On the nights of 2021 May 3 and 2021 July 9, the transits were obtained simultaneously with ESPRESSO. We reduced the data with the standard bias subtraction and flat correction procedure. The light curves were extracted using relative aperture photometry with apertures ranging from 30 to 44 pixels. The choice of reference stars and aperture size minimizes the root mean square (RMS) scatter in the out-of-transit portion. 

\begin{figure}
        \centering
        \includegraphics[width=\linewidth]{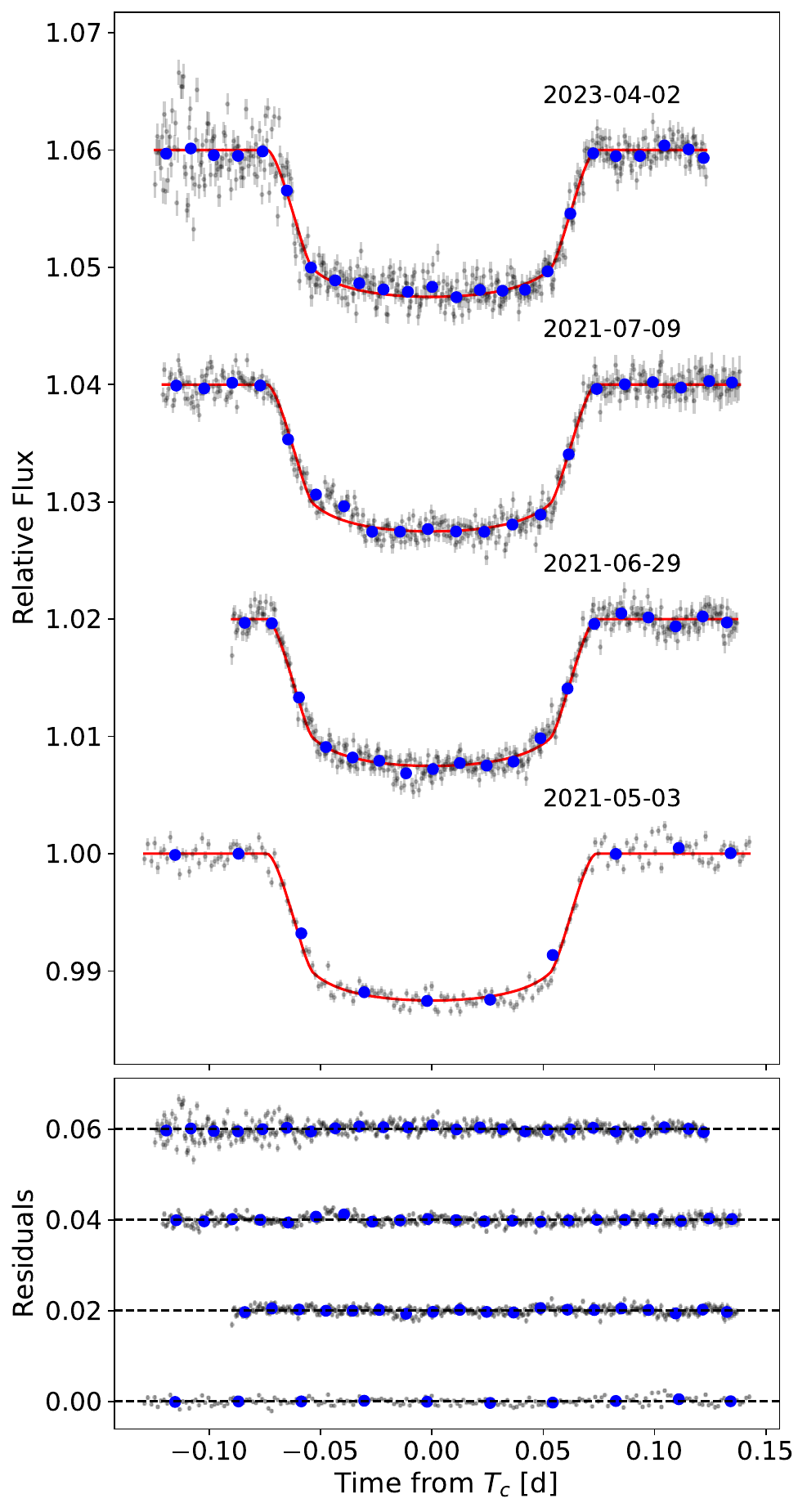}
        \caption{Four detrended EulerCam transit observations of \target with an offset of $0.02$ for visibility. {In the top panel}, the data {are} represented by the grey points and the best-fit model is represented by the red lines. {The corresponding date is displayed above each set}. The blue points correspond to the x$20$ binned data. The {bottom panel} contains the residuals for each night relative to the dashed black line, in the same order.}
        \label{fig:phot_fit}
    \end{figure}

Since the photometric data are affected by correlated noise due to observational, instrumental, or stellar effects, we used a combination of first- and second-order polynomial baseline models in different variables (time, FWHM, airmass, coordinate shifts, and sky background). The best combination of variables for each data set was established using the Bayesian information criterion (Table $\ref{tab:EulerCamphotometry}$). We detrended the light curves and derived the system parameters using \CONAN \citep{Lendl2020b}, an MCMC framework. Uniform priors were used to fit $R_p/R_\ast$, $b$, T$_0$, $P$, and T$_{14}$. The quadratic limb-darkening coefficients and their uncertainties were derived using the code LDCU\footnote{https://github.com/delinea/LDCU}  \citep{Deline2022} based on the stellar parameters derived by \citet{Hellier2019} and fitted using Gaussian priors. We also accounted for additional white noise by including a jitter term for each light curve. The set of planetary and stellar parameters we obtained and used are listed in Table \ref{tab:parameters_W178}.  The EulerCam light curves, together with their best-fit models, are presented in Fig. \ref{fig:phot_fit}.

\begin{table}
\tiny
\caption{Summary of the EulerCam light curves.}
\centering
\setlength{\tabcolsep}{10pt}\begin{tabular}{lcc}
        \hline
        \hline
        Date & Aperture (pixels) & Detrending \\
        \hline
         2021-05-03 & 44 & $p(x^2+y+FWHM^2)$ \\
         2021-07-09 & 30 & $p(t^2+x^2+y+FWHM+sky^2)$ \\
         2021-06-29 & 30 & $p(t^2+FWHM^2)$ \\
         2023-04-02 & 30 & $p(AM^2+x+y+FWHM+sky)$ \\
        \hline
    \end{tabular}
    \label{tab:EulerCamphotometry}
\begin{tablenotes}
\item
\textbf{Notes:} The notation of the baseline models for $\textit{p(j}^\textit{i}\textit{)}$ refers to a polynomial of degree \textit{i} in parameter \textit{j} (t is the time, AM is the airmass, FWHM is the stellar FWHM, and sky is the sky background). The notation of GPs refers to Gaussian processes.
\end{tablenotes}
\end{table}

\begin{table}[h!]
        \caption[]{Summary of the stellar and planetary parameters of the WASP-178 system.}
        \label{tab:parameters_W178}
        \small
        \begin{center}
                \def\arraystretch{1.25}
                \begin{tabular}{p{0.52\linewidth}p{0.33\linewidth}p{0.025\linewidth}}
                        \toprule
            \multicolumn{3}{c}{WASP-178 System parameters}  \\ \midrule
            RA$_{\rm 2000}$ & 15:09:04.89 & [4]\\
                DEC$_{\rm 2000}$ & -42:42:17.8 & [4]\\ 
                Parallax $p$ [mas] & \num{2.3119(0.0600)} & [4] \\
                Magnitude [V$_{\rm mag}$] & 9.95 & [4]\\
                Systemic velocity ($v_{\rm sys}$) [\si{\km\per\second}]    & \num{-23.908(0.007)} & [2] \\
                 \midrule
                \multicolumn{3}{c}{Stellar parameters} \\ \midrule
                Star radius ($R_\ast$) [$R_{\odot}$]    & \num{1.772(0.020)} & [3]\\
                        Star mass ($M_\ast$) [$M_{\odot}$] & \numpm{2.169}{0.083}{-0.089} & [3] \\
                        Proj. rot. velocity ($v\sin{i}$) [\si{\km\per\second}] & \numpm{10.48}{0.63}{-0.56} & [1] \\
                Age [Gyr] & \numpm{0.05}{0.06}{-0.05} & [3] \\
                Effective Temperature (T$_\mathrm{eff}$) [K] & \num{9350(150)} & [2] \\
                Surface gravity ($\log g_\star$) [$\log g_\odot$] & \num{4.35(0.15)} & [2] \\
                Metallicity ([Fe/H]) [[Fe/H]$_\odot$] & \num{0.21(0.16)} & [2] \\
                        \midrule
                        \multicolumn{3}{c}{Planetary parameters}  \\ \midrule
                        Planet radius ($R_{\rm p}) $ [$R_{\rm Jup}$]  & \num{1.796(0.022)} & [1]  \\
                        Planet mass ($M_{\rm p} $) [$M_{\rm Jup}$]  & \num{1.66(0.12)} & [2] \\
                        Eq. temperature ($T_{\rm eq}$) [$\si{\kelvin}$] & \num{2470(60)} & [2] \\
                        Density ($\rho$) [\si{\g\per\cm\cubed}] & \num{0.632(0.051)} & [1] \\
                        Surface gravity ($\log g_{\rm p}$) [cgs] & \num{3.106(0.033)} & [1] \\
                        \midrule
                        \multicolumn{3}{c}{Orbital and transit parameters} \\ \midrule
                        Transit centre time ($T_0$) [HJD (UTC)]  & \numpm{2459338.19728}{0.00007}{-0.00015} & [1] \\
                        Period ($P$) [d]   & \numpm{3.3448371}{0.0000010}{-0.0000011} & [1]  \\
                        Orbital semi-major axis ($a$) [au]   & \numpm{0.0561}{0.0005}{-0.0013}& [1] \\
                        Scaled semi-major axis ($a/R_\ast$)  & \numpm{7.00}{0.03}{-0.13}& [1]\\
                        Orbital inclination ($i$) [$^\circ$]   & \numpm{85.50}{0.09}{-0.29} & [1]  \\
                        Projected orbital obliquity ($\lambda$) [$^\circ$] & \numpm{105.7}{3.6}{-4.2} & [1] \\
                        Impact parameter ($b$) & \numpm{0.549}{0.025}{-0.008} & [1] \\
                        Eclipse duration ($T_{14}$) [h] & \numpm{3.532}{0.019}{-0.004} & [1]\\
                        Radius ratio ($R_p/R_\ast$) & \numpm{0.10716}{0.00047}{-0.00039} & [1] \\
                        RV semi-amplitude ($K_\star$)  [\si{\km\per\second}] & \num{0.139(0.009)} & [2]\\
                        Eccentricity ($e$) & \num{0} (fixed) & -  \\
                        \midrule
                        \multicolumn{3}{c}{Derived parameters} \\
                        \midrule 
                        Planetary orbital velocity ($v_{\rm orb}$)  [\si{\km\per\second}]    & \num{182.3(2.6)} & \\   
                        Approx. scale height ($H$) [\si{\km}]                    & \num{732(59)} & \\
                        Transit depth of $H$ ($\delta F / F$)  {\scriptsize [$\times 10^{-5}$]}  & \num{13.1(1.1)}  & \\

                        \bottomrule
                \end{tabular}
        \end{center}
 \tablebib{(1)~This work; (2)~\citet{Hellier2019}; (3)~\citet{Pagano2023}; (4)~\citet{Gaia2018}}
 \end{table}
    

On the day of the second ESPRESSO transit (2021 July 9), we noticed a deviation from the light-curve fitted model starting around the end of ingress. This pattern can be caused by the passage of the planet in front of a spot on the surface of the star. Although WASP-178 is an A-type star, for which no magnetic activity is expected, these stars may still present some level of magnetic activity phenomena \citep{Rackham2019}. In order to safeguard from the influence of a possible active region, the five affected in-transit spectra between $-0.054$ and $-0.030$ days from the transit centre were discarded from the analysis. Without these exposures, the S/N of the second transit transmission spectrum in the \NaI{} doublet orders drops from $402$ to $386$, while for the first transit, we reach an S/N of $456$ in these orders. 


\section{Transmission spectroscopy of \target}
\label{sec:transSpec}

    To construct the transmission spectrum, we combined the out-of-transit spectra in the stellar rest frame into the master-out spectrum, which contains the mean stellar spectrum. The master-out as well as the RM model were divided out of the in-transit data, leaving behind the signal of the planetary atmosphere. These spectra were then shifted to the planetary rest frame and combined into the transmission spectrum. We executed this process for each night separately to evaluate any variations between them. No significant difference was noted between the transmission spectra of the two nights, and they were therefore combined to form a transmission with a higher S/N. For more details of the method we followed, we refer to \citet{Wyttenbach2015} and \citet{Seidel2019}.

    It is known that the coudé train optics of ESPRESSO leave interference patterns on the data in the form of pseudo-sinusoidal waves \citep{Allart2020, Tabernero2021, Sedaghati2021}. These interference patterns are most prominent after the stellar out-of-transit spectrum is removed, and they are a conjunction of a high-frequency wave ($\sim$ 1 $\AA$ period) and a lower frequency wave ($\sim$ 30 $\AA$ period). The higher-frequency wiggles have a smaller amplitude and can be neglected as they are hidden in the noise. On the other hand, the lower-frequency wiggles present a significant effect on the transmission spectra around the \NaI doublet. We corrected for this trend by masking the sodium lines and using a cubic spline fit to the over-binned (x600) transmission spectrum of each transit. The effect of the wiggles was not noticeable in the remaining orders addressed in this work.

    \subsection{Transmission line fitting}\label{sec:line_fits}
    
\begin{figure*}
        \sidecaption
        \includegraphics[width=12cm]{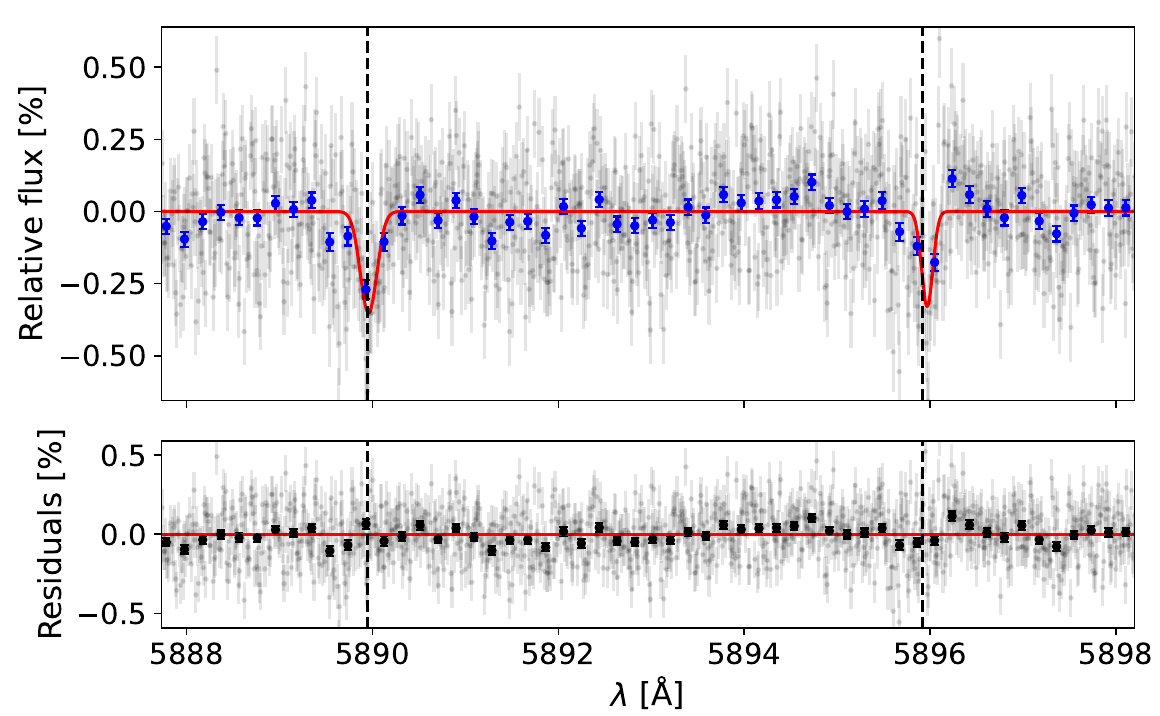}
        \caption{ESPRESSO transmission spectrum of \target on the planetary rest frame around the \NaI doublet. {\textit{Top panel:}} The grey points represent the unbinned data, the blue points represent the data binned by x20 and the Gaussian fit to the sodium doublet lines is shown by the {solid} red line. The expected line centres are marked by the dashed black lines. {\textit{Bottom panel:} The residuals of the fit are shown, with the black dots corresponding to the binned data}.}
        \label{fig:line_fit_Na}
    \end{figure*}

\begin{figure}
        \centering
        \includegraphics[width=\linewidth]{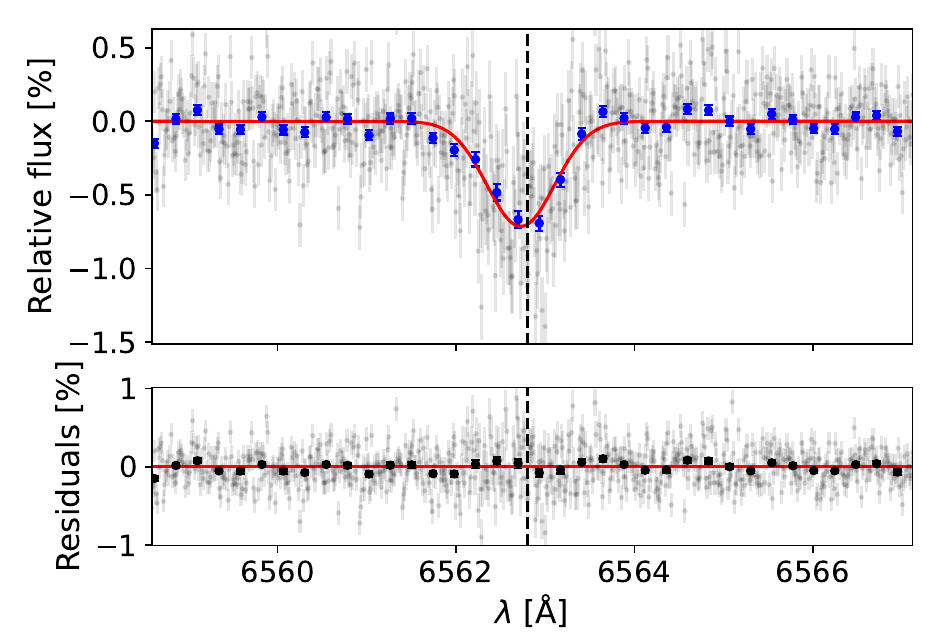}
        \caption{Same as Fig. \ref{fig:line_fit_Na} for the \halpha line.}
        \label{fig:line_fit_Ha}
    \end{figure}

\begin{figure}
        \centering
        \includegraphics[width=\linewidth]{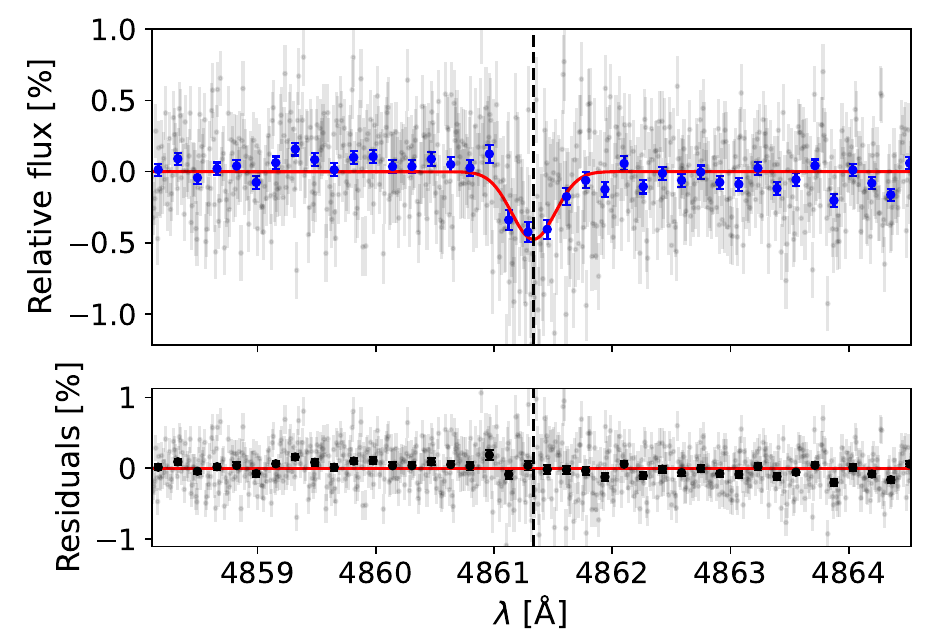}
        \caption{{Same as Fig. \ref{fig:line_fit_Na} for the \hbeta line.}}
        \label{fig:line_fit_Hb}
    \end{figure}

\begin{figure}
        \centering
        \includegraphics[width=\linewidth]{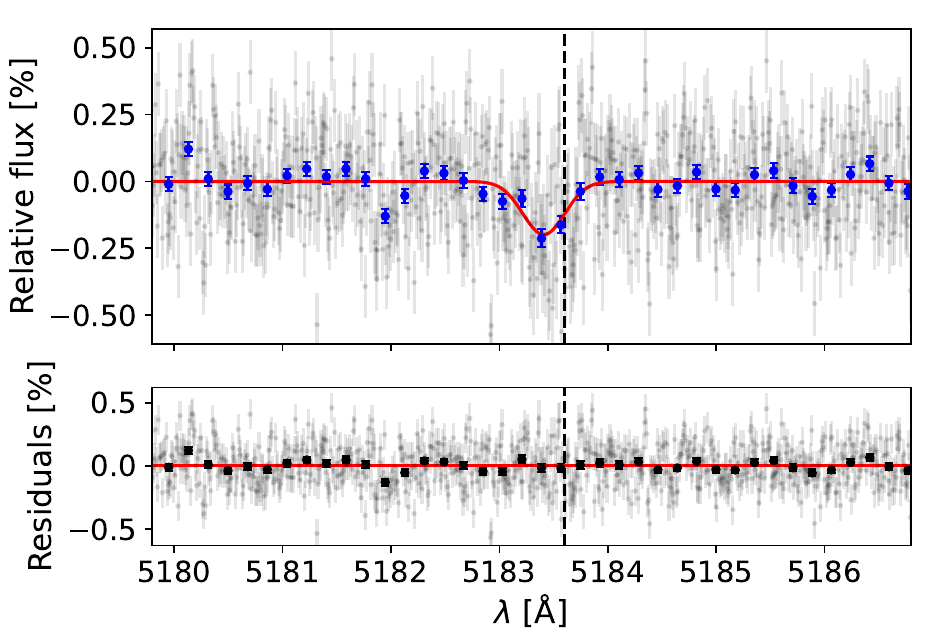}
        \caption{Same as Fig. \ref{fig:line_fit_Na} for the \MgI b1 line.}
        \label{fig:line_fit_Mg}
    \end{figure}

\begin{figure}
        \centering
        \includegraphics[width=\linewidth]{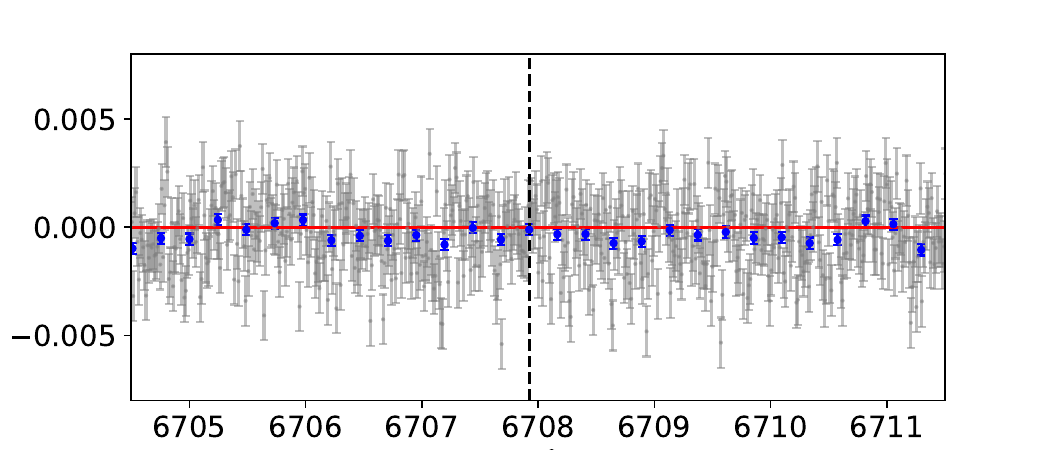}
        \caption{Same as Fig. \ref{fig:line_fit_Na} for the \LiI line. The data show no absorption signal in this region.}
        \label{fig:transmission_Li}
    \end{figure}

\begin{figure}
        \centering
        \includegraphics[width=\linewidth]{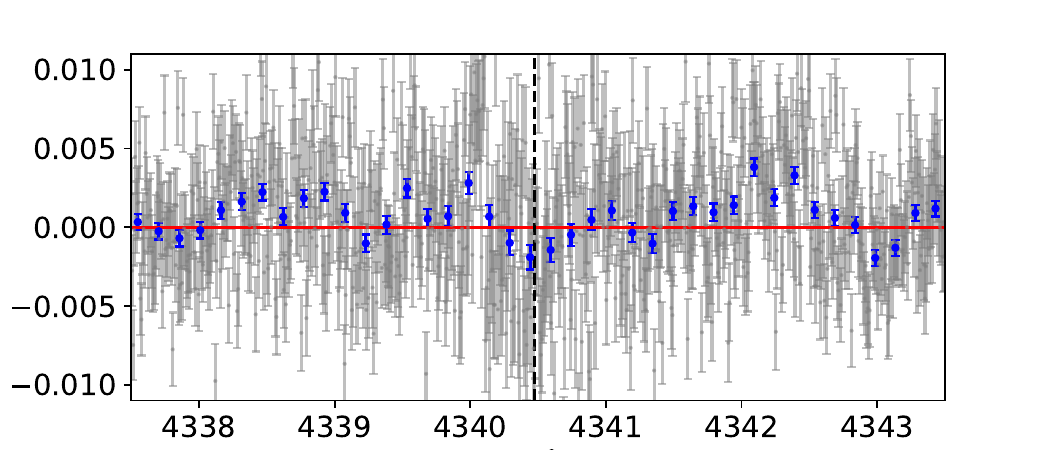}
        \caption{{Same as Fig. \ref{fig:line_fit_Na} for the \hgamma line. The data show no absorption signal in this region.}}
        \label{fig:transmission_Hg}
    \end{figure}

The transmission lines were binned and fitted with Gaussian profiles using a Bayesian MCMC approach. In Figs. \ref{fig:line_fit_Na}-\ref{fig:line_fit_Mg}, the transmission lines for the \NaI doublet, \halpha, \hbeta, and the \MgI b1 lines are shown, respectively, with their best-fit profile. {In the case of the \LiI and \hgamma lines}, no signal was identified, as we can verify from {Figs.~\ref{fig:transmission_Li}-\ref{fig:transmission_Hg}}. The resulting line fit parameters are shown in Table~\ref{tab:fit_parameters}. We used the central wavelength and FWHM of the fitted Gaussians to derive the equivalent line centre shift and broadening in terms of velocity. The detection level was computed as the ratio of the fitted line amplitude {and its positive uncertainty, obtained from the probability distribution of the parameters,} propagated with the false-positive probability. This is discussed further in Sect. \ref{sec:bootstrap}. 

We related the absorption line depths with the area covered by the respective element and translated the line profiles to equivalent heights and velocity distributions. This is shown in Figs. \ref{fig:co_added_Na}-\ref{fig:co_added_Mg}. For the \NaI doublet, we co-added the lines in velocity space with respect to their expected position. We also illustrate the instrumental broadening and the escape velocity, $v_{\rm esc}$, with height $h$ above the opaque planetary disk, calculated via Eq. \ref{eq:escape_vel}, for comparison,

\begin{equation}
    v_{esc} = \sqrt{\frac{2GM_p}{(R_p+h)}}
    \label{eq:escape_vel}.
\end{equation}

{With the detection of two of the hydrogen Balmer series lines, we are able to estimate the temperature in the thermosphere layer. This can be done through Boltzmann's equation, using the \halpha and \hbeta lines to estimate the ratio of hydrogen atoms in the second and third excited states. Since we do not have the true continuum of the planetary spectrum, we may use the equivalent widths measured from the transmission spectrum as a probe for the ratio of the electronic levels.}

\begin{table*}
        \caption{{Median values of the Gaussian fit parameters and 68\% confidence intervals from the probability distribution for the narrow-band detections.}}\vspace{-1em}
        \begin{center}
                \begin{tabular}{lllllll}
                        \toprule
                           & Amplitude [\%]  & FWHM $\left[\AA\right]$ & Centre $\left[\AA\right]$ & V$_\mathrm{FWHM}$ $\left[\kms\right]$ & Centre shift $\left[\kms\right]$ & Detection \\
                        \midrule
                        \NaI D2 &  -0.338$^{+0.053}_{-0.056}$ & 0.186$_{-0.036}^{+0.044}$ & 5889.953$_{-0.012}^{+0.013}$ & 9.5$_{-1.8}^{+2.2}$ & 0.14$_{-0.60}^{+0.65}$ & 5.5$\sigma$\\
                        \NaI D1 & -0.314$_{-0.055}^{+0.049}$ & 0.160$_{-0.034}^{+0.037}$ & 5895.966$_{-0.014}^{+0.012}$ & 8.1$_{-1.7}^{+1.8}$ & 2.18$_{-0.79}^{+0.65}$ & 5.4$\sigma$ \\
                        \halpha & -0.717$_{-0.039}^{+0.038}$ & 0.866$_{-0.044}^{+0.047}$ & 6562.724$_{-0.023}^{+0.022}$ & 39.6$_{-2.0}^{+2.1}$ & -3.5$_{-1.0}^{+1.0}$ & 13$\sigma$\\
                        \hbeta & -0.462$_{-0.062}^{+0.060}$ & 0.447$_{-0.065}^{+0.075}$ & 4861.338$_{-0.028}^{+0.032}$ & 27.6$_{-4.0}^{+4.6}$ & 0.3$_{-1.7}^{+1.9}$ & 7.1$\sigma$\\
                        \MgI b1 & -0.200$_{-0.031}^{+0.030}$  & 0.477$_{-0.081}^{+0.092}$ & 5183.427$_{-0.038}^{+0.034}$ & 27.6$_{-4.7}^{+5.3}$ & -10.2$_{-2.2}^{+2.0}$ & 4.6$\sigma$\\
                        \LiI resonant line & ... & ... & ... & ... & ... & <$3~\sigma$\\
                        \hgamma & ... & ... & ... & ... & ... & <$3~\sigma$\\
                        \bottomrule
                \end{tabular}
        \end{center}
        \vspace{-1em}
        \textit{Note:} The amplitude [\%], FWHM [$\AA$], and centre [$\AA$] were the fitted parameters, and the remaining values were derived.{\LiI and \hgamma were} not detected in the narrow-band transmission spectrum, and we therefore did not fit any profile.
        \label{tab:fit_parameters}
\end{table*}


    \subsection{Bootstrap analysis}\label{sec:bootstrap}

    We performed a bootstrap analysis through empirical Monte Carlo (EMC) methods to assess the possibility that the absorption features are induced by spurious signals or systematic errors, and quantified it through the false-positive probability. We followed the approach in \citet{Redfield2008} by comparing the results achieved from three simulated scenarios: in-in, out-out, and in-out. In all scenarios, we sampled with repetition exposures to form virtual in-transit and out-of-transit sets, each with 18 spectra. In the out-out scenario, both virtual sets contain spectra sampled from the real out-of-transit set, while the opposite was done for the in-in scenario. The third scenario was the in-out, where the virtual in-transit and out-of-transit sets contained real in-transit and out-of-transit spectra, respectively. 
    
    In each of the scenarios, we built the transmission spectrum following the steps discussed previously and computed the relative depth of the central wavelength passband in comparison to adjacent red and blue passbands. We defined central passbands with widths of 12 $\AA$ around the centre of the \NaI doublet, 10 $\AA$ around the \halpha line, and {6 $\AA$ around the \hbeta and \MgI b1 lines}. The red and blue passbands were defined with widths of 6 $\AA$ for the sodium doublet, {the \hbeta and} the \MgI b1 lines, and of 10 $\AA$ for the \halpha line.

    
    We performed 10,000 iterations of the in-in, out-out, and in-out scenarios on each night separately and built the transmission spectra. The relative depth distributions are {shown in Figs.~\ref{fig:bootstrap_narrow_1}-\ref{fig:bootstrap_narrow_2}} for {the \NaI doublet, the \halpha, \hbeta, and \MgI b1 lines}. We fitted a Gaussian profile to each of the distributions and record the in-out absorption depth, the standard deviation of the out-out distribution, $\sigma_\mathrm{out-out}$, and the false-positive probability in Table \ref{tab:bootstrap_results_narrow}. The false-positive probability was derived from $\sigma_\mathrm{out-out}$ and quantifies the possibility of a spurious detection from other than the planet's atmosphere, since the out-out scenario is composed of out-of-transit spectra alone. If the signal observed in the transmission spectrum is of planetary origin, we expect the distributions to be centred around zero, with the in-out distribution shifted. 
    
    In all cases, the in-in and out-out scenarios are centred around $0.0$ $\%$. They deviate by $0.005$ $\%$ at most, as expected. The in-out scenarios for the \NaI doublet and \halpha deviate by more than $-0.19$ $\%$ and $-0.14$ $\%$, respectively, indicating that the signals are of planetary origin. {The case for the \hbeta line shows a similar result, although with a relatively smaller deviation}. For the \MgI b1 line, we cannot make this conclusion since the in-out distributions are drifted by an order of magnitude less from $0.0$ $\%$, indicating that there is little to no absorption. {In addition, the in-out distribution on the second night shows a positive shift, which corresponds to an emission signal. This is another indication that these results are inconclusive}. We took the false-positive probability as the standard deviation of the out-out scenario, corrected to the real number of in- and out-of-transit exposures. We conclude from the bootstrap analysis that the \NaI{} doublet has a false-positive probability of $0.047~\%$ and $0.037~\%$ for the two transits, respectively. {These values are $0.056~\%$ and $0.057~\%$ for \halpha, $0.033~\%$ and $0.038~\%$ for \hbeta, and for \MgI{} they are $0.043~\%$ and $0.041~\%$.} 

    \begin{table*}
        \caption{{Results from the EMC bootstrap method for the narrow-band results.} }\vspace{-1em}
        \begin{center}
                \begin{tabular}{lllll}
                        \toprule
                            & & in-out centre [$\%$]  & $\sigma_{out-out}$ [$\%$] & False-Positive probability [$\%$] 
                           \\
                        \midrule
                        \multirow{2}{*}{\NaI} & 2021 May 03 & -0.19 & 0.065 & 0.047  \\
                        & 2021 July 09 & -0.19 & 0.055 & 0.037 \\
                        \midrule
                        \multirow{2}{*}{\halpha} & 2021 May 03 & -0.15 & 0.077 & 0.056 \\
                        & 2021 July 09 & -0.14 & 0.086 & 0.057\\
                        \midrule
                        \multirow{2}{*}{\hbeta} & 2021 May 03 & -0.040 & 0.045 & 0.033 \\
                        & 2021 July 09 & -0.036 & 0.056 & 0.038\\
                        \midrule
                        \multirow{2}{*}{\MgI b1} & 2021 May 03 & -0.011 & 0.059 & 0.043\\
                        & 2021 July 09 & 0.012 & 0.061 & 0.041 \\
                        \bottomrule
                \end{tabular}
        \end{center}
        \vspace{-1em}
        \label{tab:bootstrap_results_narrow}
\end{table*}







\section{Cross-correlation analysis}\label{sec:crosscorr}

\begin{figure*}
        \centering
        \includegraphics[width=\linewidth]{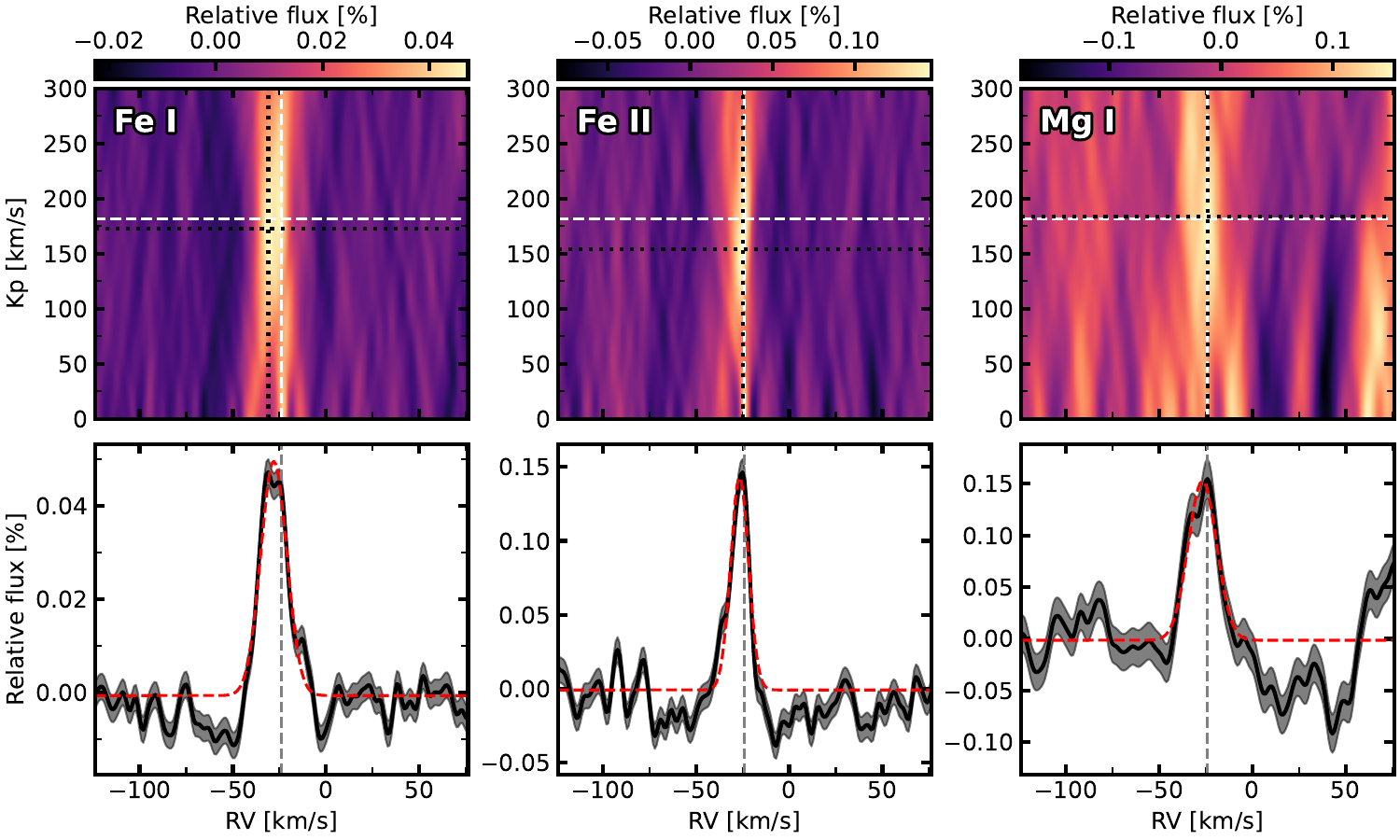}
        \includegraphics[width=\linewidth]{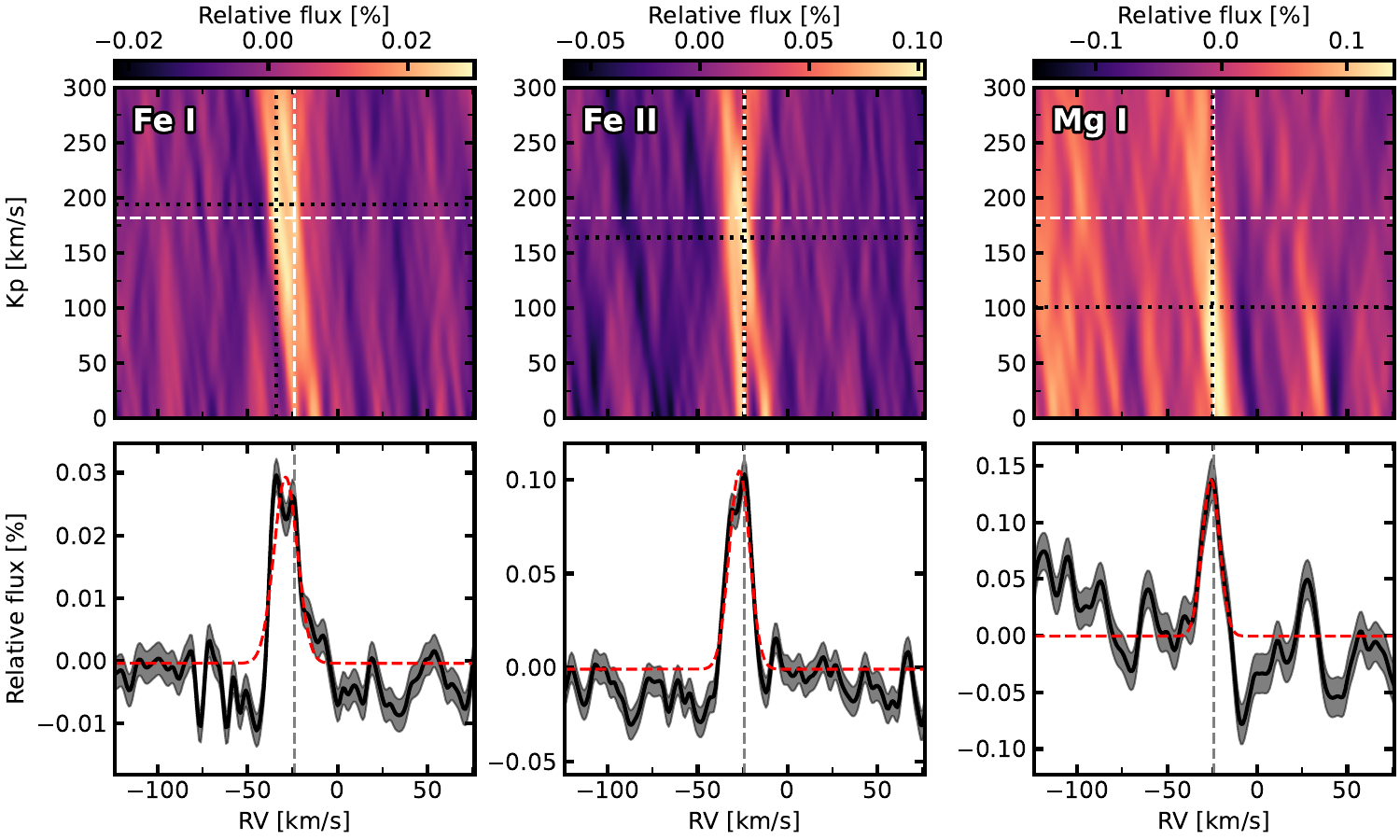}
        \caption{Cross-correlation analysis results of the first (2021 May 3, top) and second (2021 July 9, bottom) transit. From left to right, we searched for \FeI, \FeII, and \MgI. \textit{Top panel:} Expected velocities (white lines), and the detected location in the K$_p$-V$_{\rm sys}$ space (black lines). \textit{Bottom panel:} Extracted one-dimensional cross-correlation function at the detected location (horizontal black line). The dashed red lines show the best-fit Gaussian. The vertical black line shows the expected rest frame velocity, indicating blueshifts for all the detected species. }
        \label{fig:ccf}
    \end{figure*}

    To confirm the detection of \MgI and search for \FeI and \FeII, we further analysed the two transit-time series using the cross-correlation technique \citep{snellen_orbital_2010}, following the method in \citet{hoeijmakers_hot_2020}. We only included exposures with a sufficiently high S/N (see Sect.\,\ref{sec:obs}). We used \texttt{tayph}\footnote{https://github.com/Hoeijmakers/tayph} to perform the cross-correlation analysis on the S2D spectra. In particular, we corrected the spectra for telluric contamination by dividing out the best-fit telluric profiles (see Section \,\ref{sec:obs}) and then performed a velocity correction for the stellar reflex motion, leaving all spectra at a constant velocity shift that is consistent with the systemic velocity. Following \citet{hoeijmakers_hot_2020}, we applied a colour correction to the individual orders to account for variations in the broad-band continuum throughout the observations \citep[see][for a more detailed discussion]{hoeijmakers_hot_2020}, and we rejected outliers by applying an order-by-order sigma-clipping algorithm with a width of 40 px. This algorithm computes a running median absolute deviation over sub-bands, and any value that deviated by more than 5$\,\sigma$ from the running median in any given band was masked out, thus rejected, and interpolated. Additionally, we manually flagged columns in the data where the telluric correction left systematic residuals, in particular, in the region of deep O$_2$ bands, and we rejected the reddest order because \MgI, \FeI, and \FeII are not expected to absorb significantly at the reddest orders. The rejection of outliers and manual masking affected 4.31\% and 13.53\% of the pixels in the time series of 2021 May 3 and 2021 July 9, respectively. Using cross-correlation templates from \citet{kitzmann_mantis_2023} at a temperature of \SI{2500}{\kelvin} (neutrals) and \SI{4000}{\kelvin} (ion) , we searched for \MgI, \FeI and \FeII. To fit for the Rossiter-McLaughlin effect, we constructed an empirical model based on the two-dimensional cross-correlation maps for each species individually, fitting a Gaussian with two components similar to \citet{Prinoth2022} and dividing it out (see Fig. \ref{fig:RM_ccf}).
    Any residual broad-band variation was removed by a high-pass filter with a width of \SI{100}{\km\per\second}. Finally, we converted the two-dimensional cross-correlation maps into K$_p$-V$_{\rm sys}$ maps (see \citet{hoeijmakers_hot_2020} and \citet{Prinoth2022} for an extensive discussion). Using the average S/N in Table\,\ref{tab:observations} as a weight, we combined the two K$_p$-V$_{\rm sys}$ maps. The results of the cross-correlation analysis for \MgI, \FeI, and \FeII are shown in Fig.\,\ref{fig:ccf}. At the location of the peak, we extracted the one-dimensional cross-correlation function and fitted a Gaussian to determine the amplitude, centre, and width of the absorption. The fitted parameters are shown in Table \ref{tab:CCF_fit_parameters}, along with the shift between the literature V$_{\rm sys}$ value, which corresponds to the planetary rest frame. The detection level is obtained as the ratio of the fitted signal amplitude and its uncertainty.

    To assess whether the signal uniquely stems from in-transit exposures, we conducted a bootstrapping analysis, following the method outlined in \citet{hoeijmakers_hot_2020} in Appendix C.1. This assessment is similar to the approach detailed in Sect. \ref{sec:bootstrap}, but it is executed in velocity space rather than wavelength space. This adaptation takes into account the fact that we considered the average of all lines within the spectrograph wavelength and not single absorption lines. We conducted the bootstrap analysis for each night and each species individually (see \ref{app:bootstrap}). 
    


\begin{table*}
        \caption{Median values of the Gaussian fit parameters and 68\% confidence intervals from the probability distribution for the cross-correlation detections of each night.}\vspace{-1em}
        \begin{center}
                \begin{tabular}{llllllll}
                        \toprule
                            & {Date} & Amplitude $\left[\%\right]$  & FWHM $\left[\kms\right]$ & Peak RV $\left[\kms\right]$ & Peak RV-V$_{sys}$ shift $\left[\kms\right]$ & Detection \\
                        \midrule
                        \multirow{2}{*}{\FeI} & 2021 May 03 & \num{0.0502(0.0015)} & \num{16.53(0.57)} & \num{-28.10(0.24)} & \num{-4.19(0.24)} & 12 \\
                        & 2021 July 09 & \num{0.0297(0.0016)} & \num{15.36(0.94)} & \num{-28.83(0.40)} & \num{-4.93(0.40)} & 10\\
                        \midrule
                        \multirow{2}{*}{\FeII} & 2021 May 03 & \num{0.1441(0.0058)} & \num{11.32(0.52)} & \num{-26.61(0.22)} & \num{-2.70(0.22)} & 11 \\
                        & 2021 July 09 & \num{0.1056(0.0051)} & \num{13.33(0.74)} & \num{-26.72(0.31)} & \num{-2.81(0.31)} & 8.4 \\
                        \midrule
                        \multirow{2}{*}{\MgI} & 2021 May 03 & \num{0.154(0.011)} & \num{17.9(1.5)} & \num{-26.73(0.64)} & \num{-2.82(0.64)} & 7.8\\
                        & 2021 July 09 & \num{0.138(0.017)} & \num{11.4(1.6)} & \num{-25.74(0.67)} & \num{-1.83(0.67)} & 5.8\\
                        \bottomrule
                \end{tabular}
        \end{center}
        \vspace{-1em}
        \textit{Note:} All parameters shown were obtained from the fit to the CCF signal, except for the peak shift (RV-V$_{sys}$) and the detection level.
        \label{tab:CCF_fit_parameters}
\end{table*}

\begin{table*}
        \caption{Results from the EMC bootstrap method for the cross-correlation results. }\vspace{-1em}
        \begin{center}
                \begin{tabular}{lllll}
                        \toprule
                            & & in-out centre [$\%$]  & $\sigma_{out-out}$ [$\%$] & False-Positive probability [$\%$] 
                           \\
                        \midrule
                        \multirow{2}{*}{\FeI} & 2021 May 03 & -0.030 & 0.0052 & 0.0038  \\
                        & 2021 July 09 & -0.023 & 0.0040 & 0.0025 \\
                        \midrule
                        \multirow{2}{*}{\FeII} & 2021 May 03 & -0.100 & 0.017 & 0.012 \\
                        & 2021 July 09 & -0.096 & 0.018 & 0.011\\
                        \midrule
                        \multirow{2}{*}{\MgI} & 2021 May 03 & -0.091 & 0.022 & 0.016\\
                        & 2021 July 09 & -0.052 & 0.027 & 0.017 \\
                        \bottomrule
                \end{tabular}
        \end{center}
        \vspace{-1em}
        \label{tab:bootstrap_results_cross}
\end{table*}

\section{Discussion and conclusions}
    \label{sec:diss}

    The results from the narrow-band transmission spectroscopy revealed sodium via the detection of the \NaI D2 ($5.5\,\sigma$) and D1 ($5.4\,\sigma$) lines and {hydrogen via the \halpha ($13\,\sigma$) and \hbeta ($7.1\,\sigma$) lines}, with a tentative detection of the \MgI b1 ($4.6\,\sigma$) line. According to the analysis described in Sect. \ref{sec:line_fits}, we observe a certain difference in behaviour between the detected elements. The \NaI{} D2 line does not appear to be shifted, with a centre velocity of \num{0.14(0.65)} \kms, while the D1 line shows a very slight \num{2.18(0.79)} \kms redshift (significant below $3\,\sigma$). Both lines are significantly broader than the {instrumental resolution of $2~$\kms}, with an FWHM of \num{9.5(2.2)} for the D2 line and \num{8.1(1.8)}$~$\kms for the D1 line. We observe a blueshift of \num{-3.5(1.0)}$~$\kms (significant at $3.5\,\sigma$) with a broadening of \num{39.6(2.1)} \kms for \halpha, {while \hbeta shows no shift and is broadened by \num{27.6(4.6)} \kms}. These values are far above the instrumental resolution, which could serve as an indication for possible atmospheric escape. From inspecting Figs. \ref{fig:co_added_Na}-\ref{fig:co_added_Mg}, we note that it is possible for hydrogen higher up in the atmosphere to have speeds exceeding the escape velocity, while for sodium and magnesium, this scenario is far less likely. The \MgI{} b1 line studied in the narrow band shows a considerable blueshift of \num{-10.2(2.2)} \kms. The line FWHM corresponds to a broadening of \num{27.6(5.3)} \kms, which, similarly to \halpha, could be caused by atmospheric dynamics under the caveat of our bootstrapping results in the narrow band. As this line probes the atmosphere in a lower region, where the escape of magnesium is less likely, this result points towards the existence of strong winds in the terminator. Similar atmospheric dynamics are indicated by the broadened absorption features on the hot dayside in emission, which were verified on CRIRES+ data in the {near infrared (NIR)} \citep{Cont2024}. 

    {We additionally report the non-detection of the \LiI resonance line and the \hgamma line}. The non-detection of \LiI is curious because this trace element has been detected in the atmospheres of similar UHJ, such as WASP-121 b \citep{Borsa2021} and WASP-76 b \citep{Tabernero2021}. A detection of \LiI should be made easier by the absence of \LiI absorption from the star, yielding a higher S/N in the wavelength range around the line as well as avoiding the RM effect. This might be due to insufficient S/N in our data. We note that \LiI absorption was only detected on 4-UT data and not detected on 1-UT data of WASP-121 b for this reason \citep{Borsa2021}. Further observations might help clarify the existence of \LiI in the atmosphere of \target.

    When the data were analysed through the cross-correlation method, we detected \FeI and \FeII in the atmosphere of \target with a significance of $12\,\sigma$ and $10\,\sigma$, respectively, for the first night, and with 11$\,\sigma$ and $8.4\,\sigma$ for the second night. The bootstrap analyses of these two species showed that the signal uniquely and uniformly originated from the in-transit exposures (see Appendix\,\ref{app:bootstrap}). Similarly, \MgI exhibited a signal in the K$_{\rm p}$-V${\rm sys}$ map with a significance of $7.8\,\sigma$ and $5.8\,\sigma$ for the two transits, although we note that for the second night, the location of the \MgI absorption is observed at a significantly lower K$_{\rm p}$.
    All signals appear to be blueshifted with respect to the expected rest frame velocity by $-4.19 \pm 0.24$ \kms (\FeI), $-2.70 \pm 0.22$ \kms (\FeII), and $-1.83 \pm 0.67$ \kms (\MgI) at least, and they are broadened to a FWHM of $15.36 \pm 0.94$ \kms (\FeI), $11.32 \pm 0.52$ \kms (\FeII), and  $17.9 \pm 1.5$ \kms (\MgI). These results indicate atmospheric dynamics in the more central to lower layers, where \FeI sits. Between the detections of \NaI, \FeI, \FeII, and \MgI, we observe decreasing broadening and blueshift with height, which might be an indication that the lower layers have stronger dynamics and winds because atmospheric escape is less likely for these elements. However, the disparity between the \MgI signature on both nights through cross-correlation, and with regard to the detection in narrow-band, leaves room for doubting the existence and concrete behaviour of magnesium in the atmosphere of \target.



{The ratio of the depths of \halpha and \hbeta is \num{1.55(0.22)}, which is similar to results obtained for distinct UHJ, such as WASP-33 b \citep{Cauley2021} and MASCARA-2 b \citep{CasasayasBarris2019}. Following these studies, \hgamma is expected to have a depth of $0.56$ times that of \halpha, corresponding to $-0.40~\%$, which coincides with the standard deviation of the continuum near the line centre. The noise level verified around the \hgamma line core can mask a potential detection, reinforcing the need for observations with a higher S/N. The application of Boltzmann's equation on \halpha and \hbeta yielded a temperature of \num{4875(1750)} K in the thermosphere of \target, taking into consideration that it is determined using the direct ratio of the two Balmer lines, which only serves as a rough estimate. This value agrees with the pressure-temperature profile obtained in \citet{Lothringer2022}.}

A previous study conducted with HST/WFC3/UVIS data \citep{Lothringer2022} in the NUV presented the absence of \FeI in the atmosphere of \target and found that the absorption in this band could be explained by SiO at solar abundance or \MgI{} + \FeII at super-solar abundance. The non-detection of \FeII and \MgII at higher resolution in STIS E230M data led to the conclusion that the atmosphere is likely dominated by SiO. Our detections of \MgI{}, \FeI, and \FeII in cross-correlation suggest that the most likely scenario would be a \MgI{}- and \FeII-dominated absorption in the NUV, possibly containing contributions from SiO as well in order to justify the absorption depths observed in this band. Further space-based observations could help to determine the relative abundances of these and other elements and to better interpret these results.

Strong evidence of \water{} and CO was also found on the dayside of \target through emission spectroscopy with CRIRES+ \citep{Cont2024}. As a possible scenario to explain both our findings, we propose that \water{} vapour on the nightside dissociates due to the extremely high temperature and irradiation on the dayside, allowing for the detection of hydrogen in the higher layers of the atmosphere via the \halpha{} line. Near the terminator, recombination can occur, forming once again \water{}. Extending this cycle to \Htwo{} would agree with results reported in recent studies \citep{Bell2018, Tan2019, Mansfield2020, Helling2021, Helling2023, Pagano2023}, suggesting that at the high temperatures of UHJ, such as \target, dissociation and recombination of \Htwo{} can occur, which might cause winds of ionised particles on the dayside that are driven by the magnetic field of the planet. The presence of such winds would also support the hypothesis that UHJs with high irradiation temperatures present an efficient recirculation from the day- to the nightside \citep{Zhang2018}, as was observed for instance in \citet{Seidel2021} and \citet{Seidel2023}. In a separate study, \target was found to have a very high recirculation efficiency \citep{Pagano2023}, which according to global circulation models (GCM) from \citet{Kataria2016} can be associated with the existence of zonal winds. Our results from the transmission spectroscopy show only a slight blueshift in the upper or middle layers of the atmosphere, indicating a lack of strong zonal winds in these layers. This seems to contradict previous predictions because the observed high recirculation efficiency appears to be dissociated with the presence of zonal winds.

\target is an interesting case study for atmospheric dynamics and the relation between winds and the efficiency of recirculation in UHJ. We suggest follow-up observations and investigation of this system through the use of both space-based instruments such as JWST and ground-based instruments, in order to better constrain the atmospheric composition of \target. Additionally, the application of circulation models would allow for a better constraint of the undergoing atmospheric dynamics, linking it to other hot Jupiter and UHJ observations.

\begin{acknowledgements}
{We thank the anonymous referee for their comments which have improved the manuscript.} We thank D. Cont for his fruitful discussion on this target and for sharing a draft of his parallel work on the dayside of this planet. The authors acknowledge the ESPRESSO project team for its effort and dedication in building the ESPRESSO instrument. This work relied on observations collected at the European Southern Observatory. This work has been carried out within the framework of the National Centre of Competence in Research PlanetS supported by the Swiss National Science Foundation under grants 51NF40\_182901 and 51NF40\_205606. The authors acknowledge the financial support of the SNSF. We acknowledge financial support from the Agencia Estatal de Investigaci\'on of the Ministerio de Ciencia e Innovaci\'on MCIN/AEI/10.13039/501100011033 and the ERDF “A way of making Europe” through project PID2021-125627OB-C32, from the Centre of Excellence “Severo Ochoa” award to the Instituto de Astrofisica de Canarias, and from The Fund of the Walter Gyllenberg Foundation. {This project has received funding from the Swiss National Science Foundation (SNSF) for project 200021\_200726.}  R. A. is a Trottier Postdoctoral Fellow and acknowledges support from the Trottier Family Foundation. This work was supported in part through a grant from the Fonds de Recherche du Qu\'ebec - Nature et Technologies (FRQNT). This work was funded by the Institut Trottier de Recherche sur les Exoplane\`etes (iREx). JIGH, ASM, RR and CAP acknowledge financial support from the Spanish Ministry of Science and Innovation (MICINN) project PID2020-117493GB-I00. {This work was financed by Portuguese funds through FCT (Funda\c{c}\~ao para a Ci\^encia e a Tecnologia) in the framework of the project 2022.04048.PTDC (Phi in the Sky, DOI 10.54499/2022.04048.PTDC)}. CJM also acknowledges FCT and POCH/FSE (EC) support through Investigador FCT Contract 2021.01214.CEECIND/CP1658/CT0001 (DOI 10.54499/2021.01214.CEECIND/CP1658/CT0001). FPE and CLO would like to acknowledge the Swiss National Science Foundation (SNSF) for supporting research with ESPRESSO through the SNSF grants nr. 140649, 152721, 166227, 184618 and 215190. The ESPRESSO Instrument Project was partially funded through SNSF’s FLARE Programme for large infrastructures. J.L.-B. was partly funded by grants LCF/BQ/PI20/11760023, Ram\'on y Cajal fellowship with code RYC2021-031640-I, and the Spanish MCIN/AEI/10.13039/501100011033 grant PID2019-107061GB-C61. E. E-B. acknowledges financial support from the European Union and the State Agency of Investigation of the Spanish Ministry of Science and Innovation (MICINN) under the grant PRE2020-093107 of the Pre-Doc Program for the Training of Doctors (FPI-SO) through FSE funds. A.R.C.S. acknowledges support from the Funda\c{c}\~ao para a Ci\^encia e a Tecnologia (FCT) and POCH/FSE through the fellowship 2021.07856.BD and the research grants UIDB/04434/2020 and UIDP/04434/2020. This work has been carried out within the framework of the NCCR PlanetS supported by the Swiss National Science Foundation under grant 51NF40\_205606. This work was co-funded by the European Union (ERC, FIERCE, 101052347). Views and opinions expressed are however those of the authors only and do not necessarily reflect those of the European Union or the European Research Council. Neither the European Union nor the granting authority can be held responsible for them. {This work was supported by FCT - Funda\c{c}\~ao para a Ci\^encia e a Tecnologia through national funds and by FEDER through COMPETE2020 - Programa Operacional Competitividade e Internacionaliza\c{c}\~ao by these grants: UIDB/04434/2020; UIDP/04434/2020.} E.H.-C. acknowledges support from grant PRE2020-094770 under project PID2019-109522GB-C51 funded by the Spanish Ministry of Science and Innovation / State Agency of Research, MCIN/AEI/10.13039/501100011033, and by ‘ERDF, A way of making Europe’. {Y.C.D. and J.V.S. acknowledge support from ESO through an SSDF studentship.} {MRZO acknowledges financial support from the Spanish Ministry for Science, Innovation and Universities via project PID2022-137241NB-C42. ML acknowledges support of the Swiss National Science Foundation under grant number PCEFP2\_194576.}
\end{acknowledgements}

\bibliographystyle{aa}
\bibliography{biblio.bib}

\begin{appendix}

    \section{Velocity-height narrow-band absorptions}
    \begin{figure}[h]
        \centering
        \includegraphics[width=\linewidth]{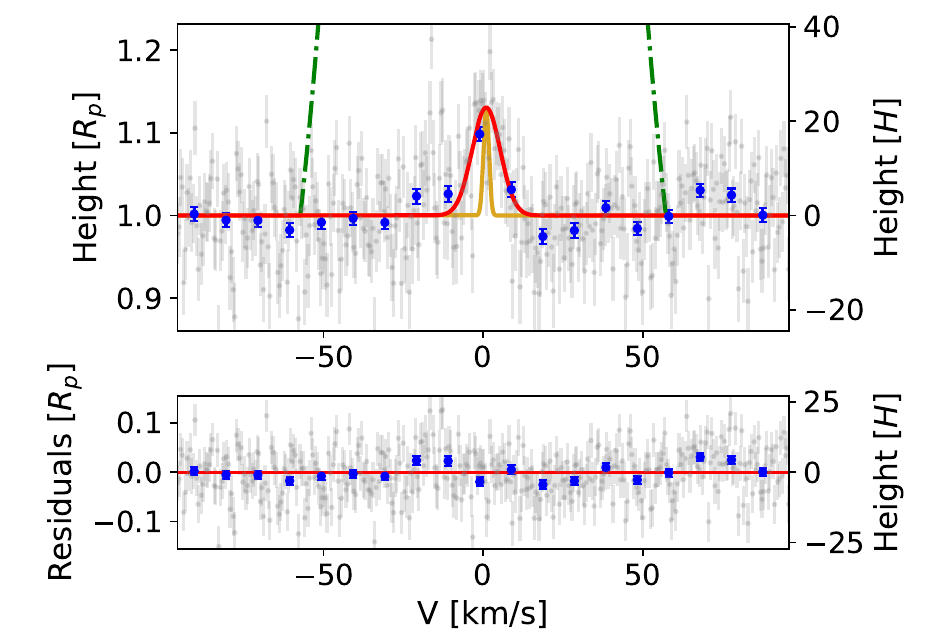}
        \caption{Co-added \NaI doublet line as a function of the velocity shift from the lines' {centre} wavelength. The absorption depth was converted to an equivalent radius. \textit{Top panel:} The data is shown by the grey points and the x20 binned data by the blue points. The data was fitted with a Gaussian, shown by the {solid} red line, while the golden line shows the ESPRESSO line spread function ({FWHM = $2$ \kms}) and the green dot-dashed line represents the escape velocity. {\textit{Bottom panel:} The residuals are presented with the same representation as the top panel.}}
        \label{fig:co_added_Na}
    \end{figure}
    \begin{figure}[h]
        \centering
        \includegraphics[width=\linewidth]{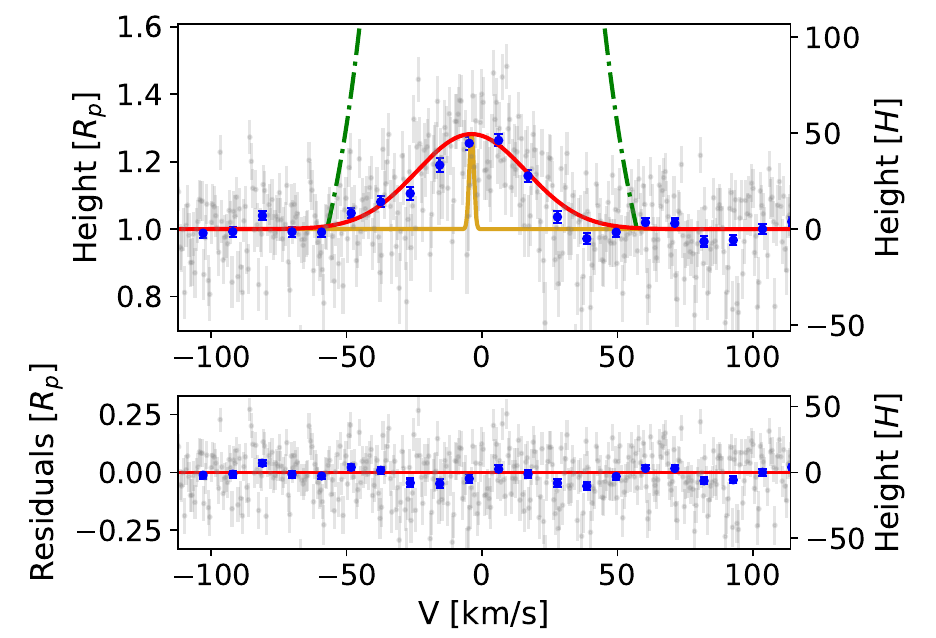}
        \caption{Same as Fig. \ref{fig:co_added_Na} for the \halpha line.}
        \label{fig:co_added_Ha}
    \end{figure}
    \begin{figure}[h]
        \centering
        \includegraphics[width=\linewidth]{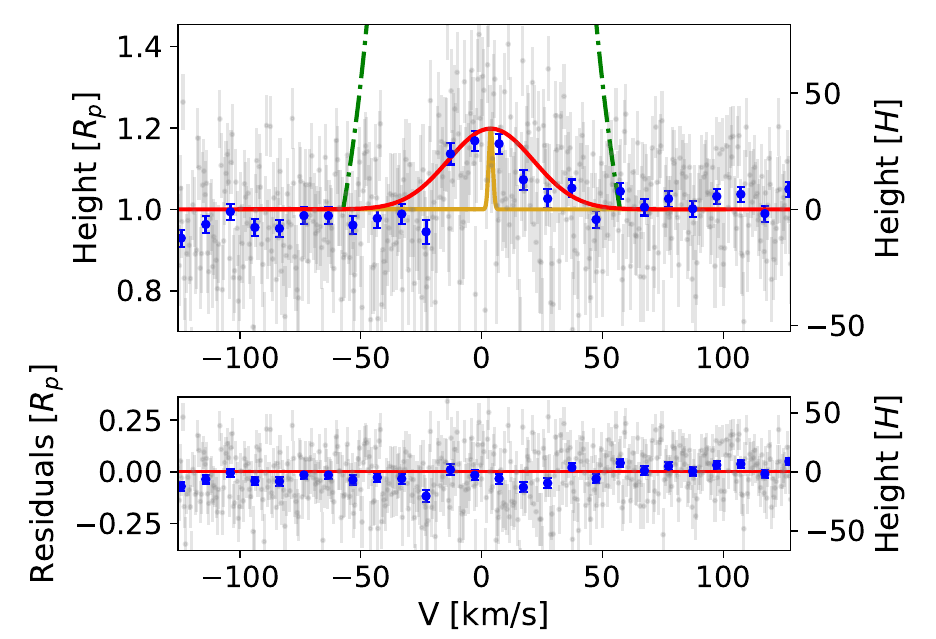}
        \caption{{Same as Fig. \ref{fig:co_added_Na} for the \hbeta line.}}
        \label{fig:co_added_Hb}
    \end{figure}
    \begin{figure}[h]
        \centering
        \includegraphics[width=\linewidth]{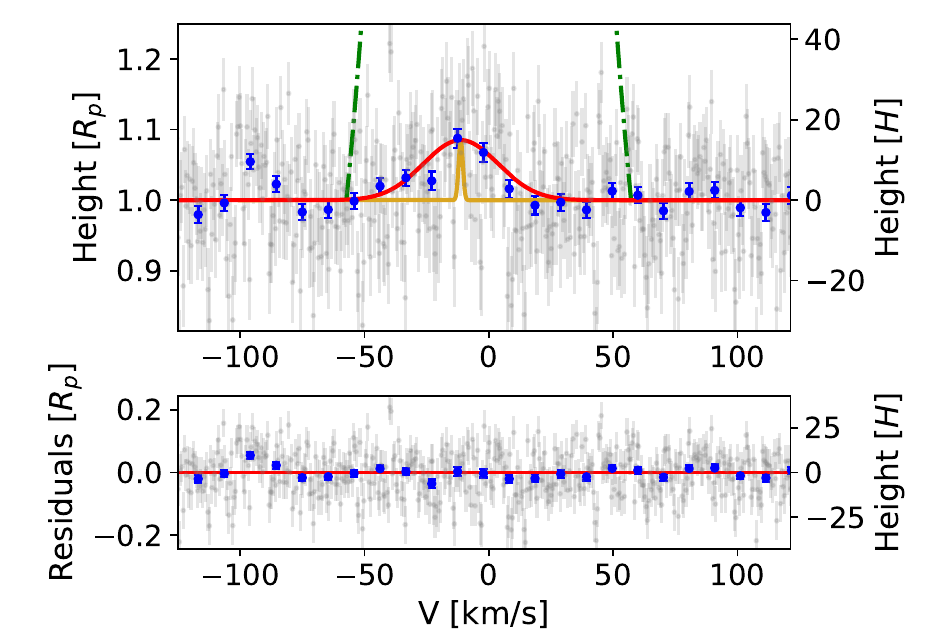}
        \caption{Same as Fig. \ref{fig:co_added_Na} for the \MgI b1 line.}
        \label{fig:co_added_Mg}
    \end{figure}

\clearpage
\onecolumn
    \section{Bootstrap analysis for the narrow band}
    \label{app:bootstrap_narrow}

    \begin{figure*}[h!]
        \centering
        \includegraphics[width=0.9\linewidth]{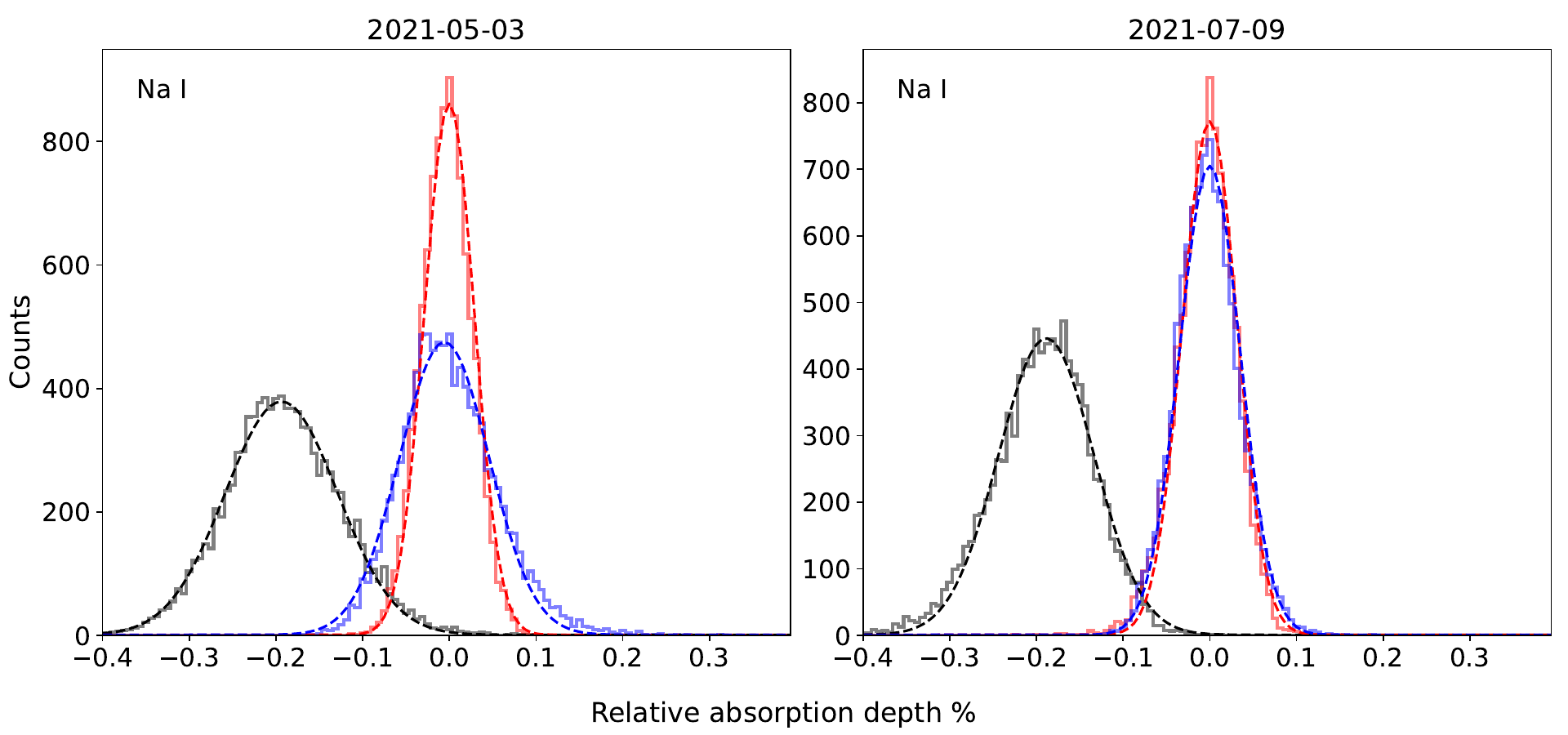}
        \includegraphics[width=0.9\linewidth]{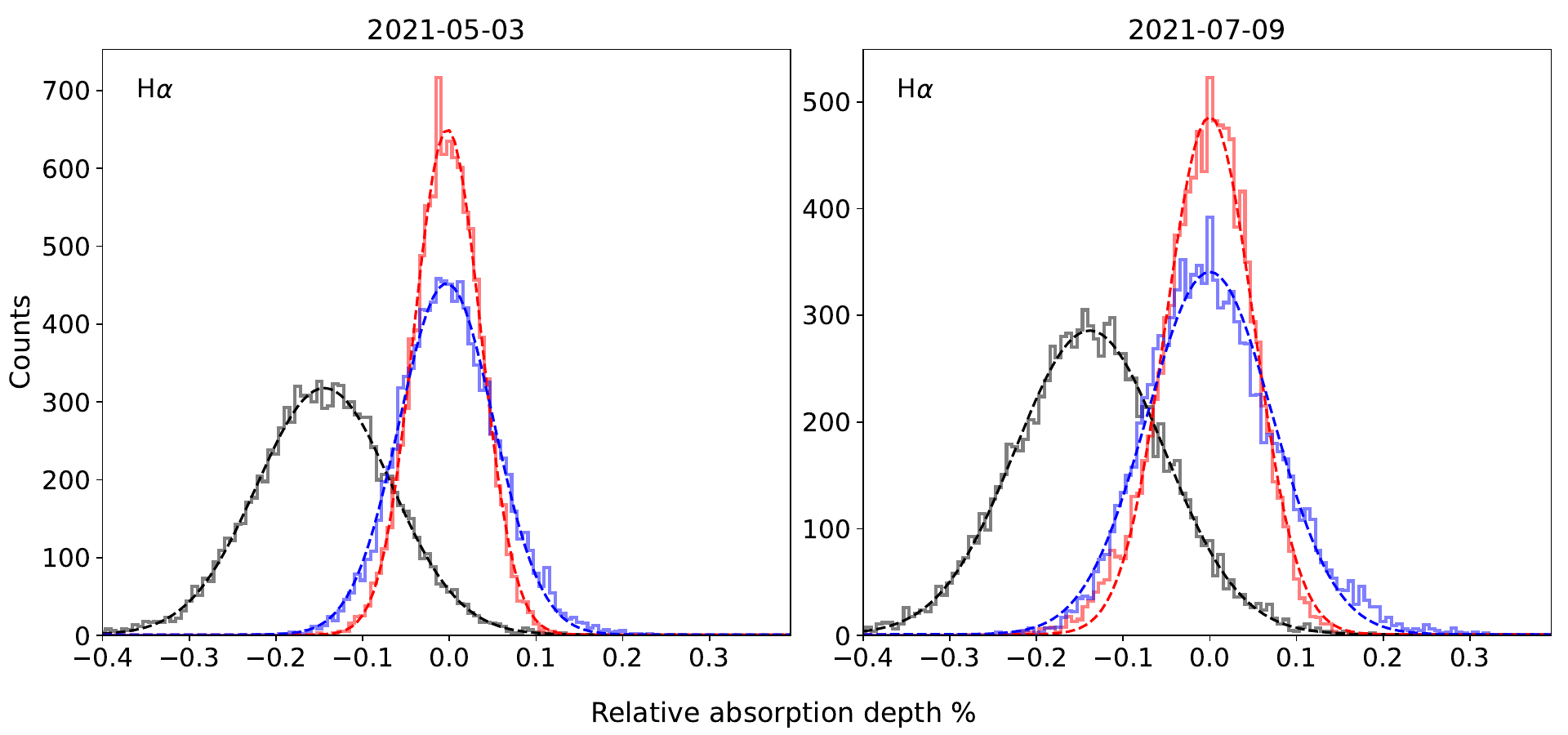}
        \caption{Distributions of the empirical Monte Carlo analysis for {the} 12 $\AA$ passband on the \NaI doublet (top panels) and 10 $\AA$ on \halpha (bottom panels), for each of the ESPRESSO transits. {The results from the first and second night are displayed in the left and right panels, respectively}. The distributions representing the in-in (red {bins}) and out-out (blue {bins}) scenarios are centred at zero, corresponding to non-detections, as expected. The in-out distributions (black {bins}) for \NaI and \halpha deviate from zero in both transits, indicating a detection of planetary origin. {The Gaussian fits to the histograms are shown as dashed lines of the respective colour.}}
        \label{fig:bootstrap_narrow_1}
    \end{figure*}
    \begin{figure*}[h!]
        \centering
        \includegraphics[width=0.9\linewidth]{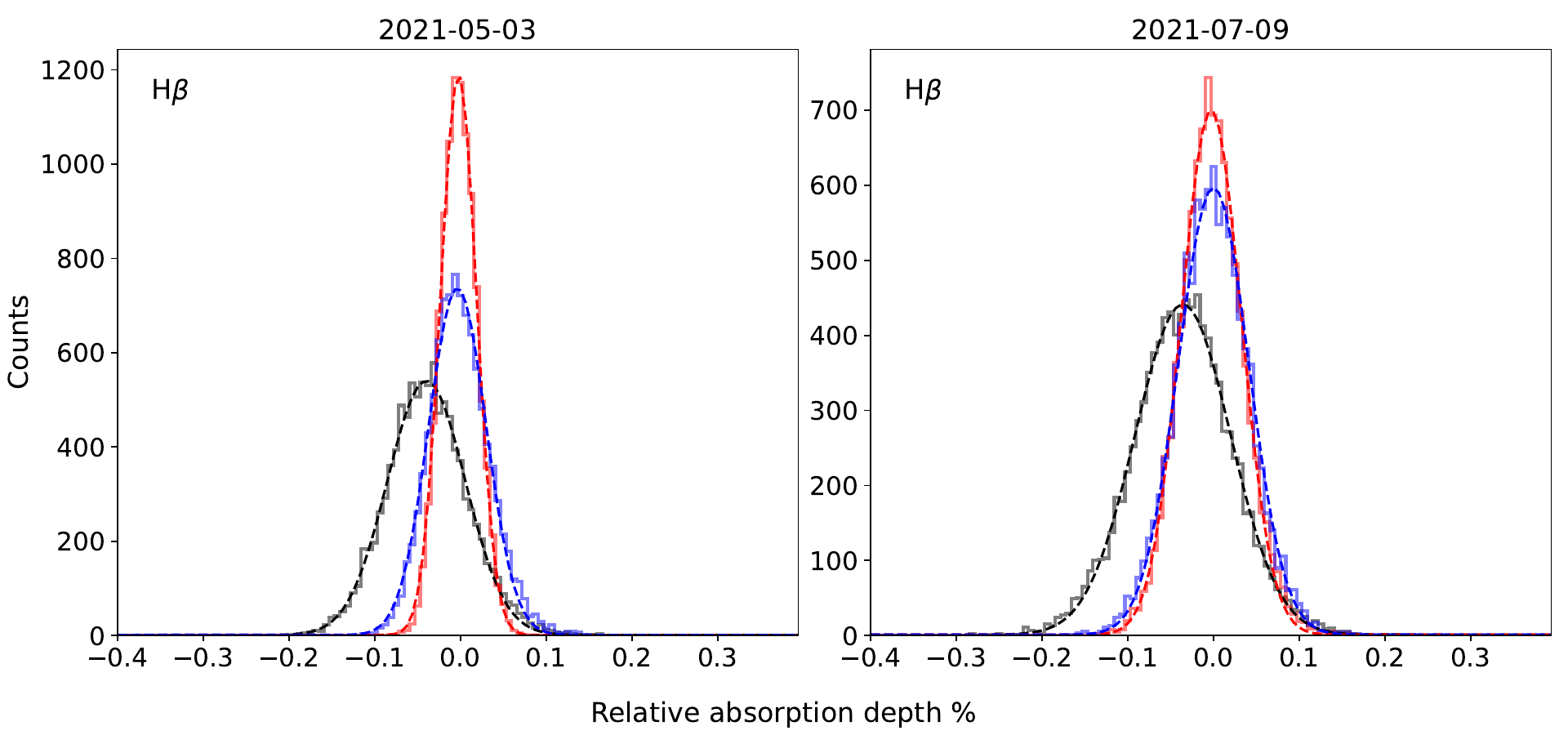}
        \includegraphics[width=0.9\linewidth]{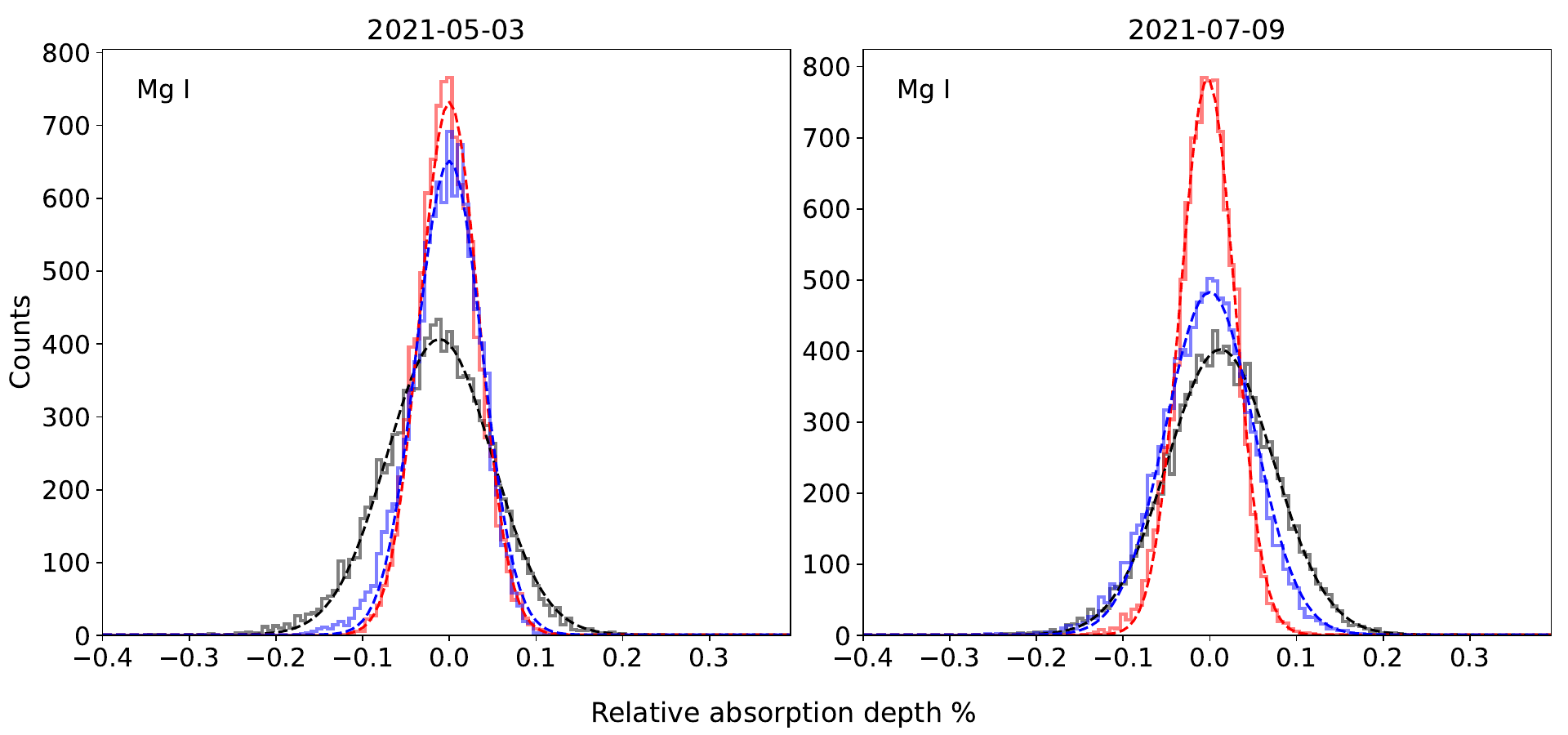}
        \caption{{Same as Fig. \ref{fig:bootstrap_narrow_1} for the 6 $\AA$ passband on the \hbeta line (top panels) and on the \MgI b1 line (bottom panels). For \hbeta, both nights deviate slightly, which hints at a detection, although not as clearly as \NaI and \halpha. In the case of \MgI, the in-out sits close to zero, indicating a lack of planetary signal.}}
        \label{fig:bootstrap_narrow_2}
    \end{figure*}

\clearpage
    \section{Bootstrap analysis for the cross-correlation}
    \label{app:bootstrap}
    \begin{figure}[h!]
        \includegraphics[width=6cm]{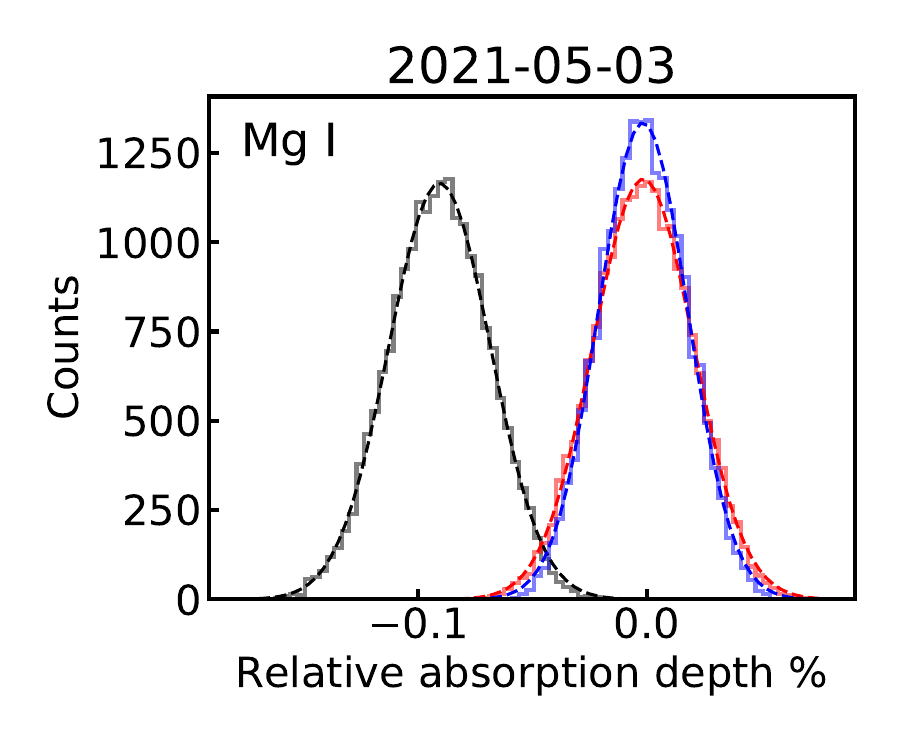}
        \includegraphics[width=6cm]{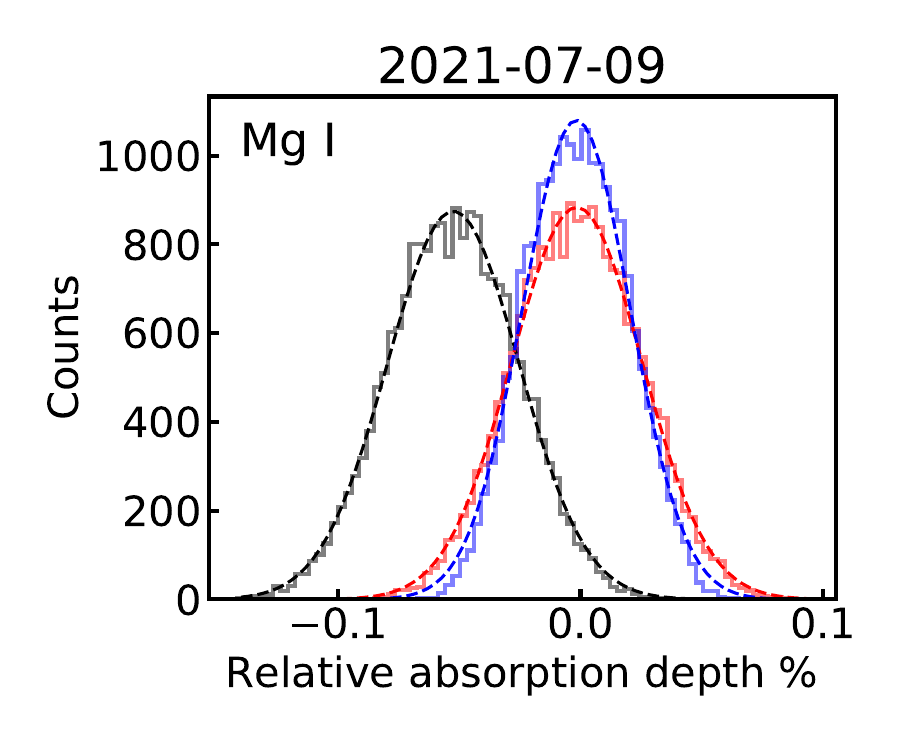}\\
        \includegraphics[width=6cm]{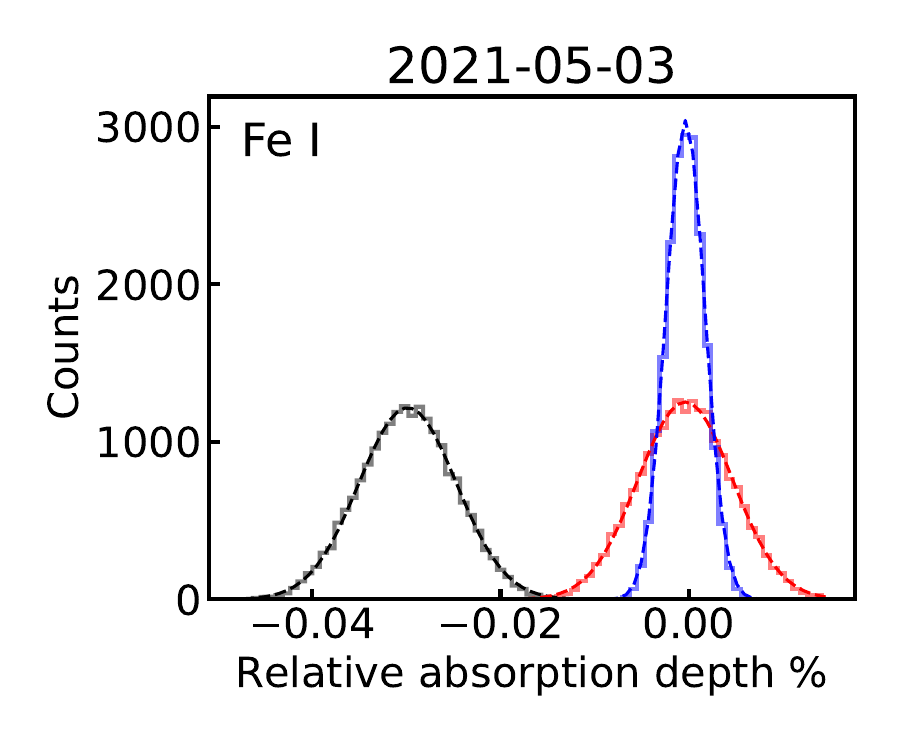}
        \includegraphics[width=6cm]{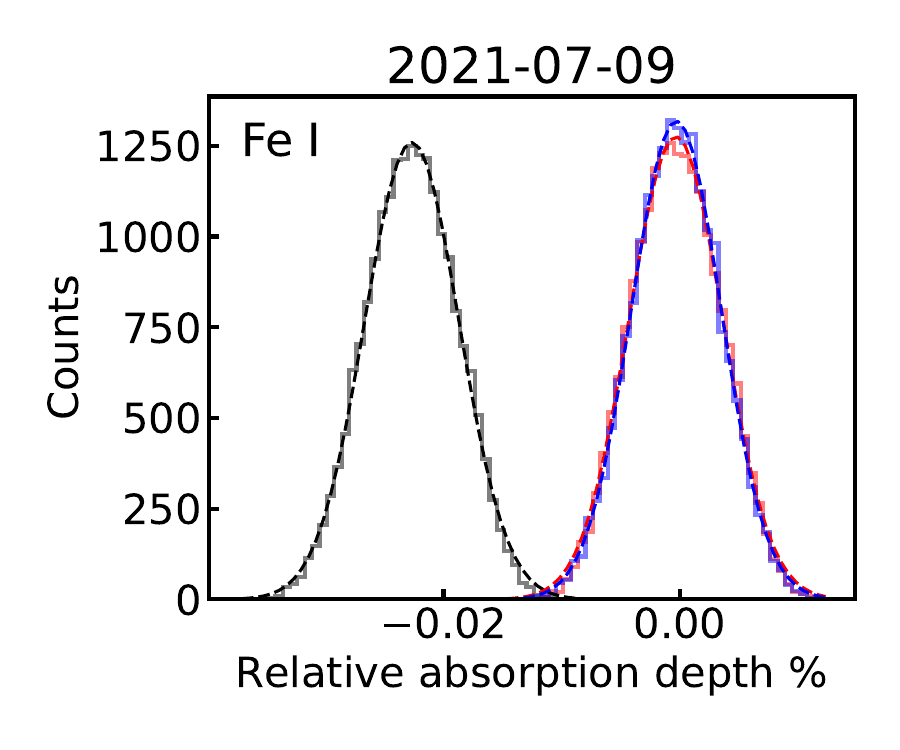}\\
        \sidecaption
        \includegraphics[width=6cm]{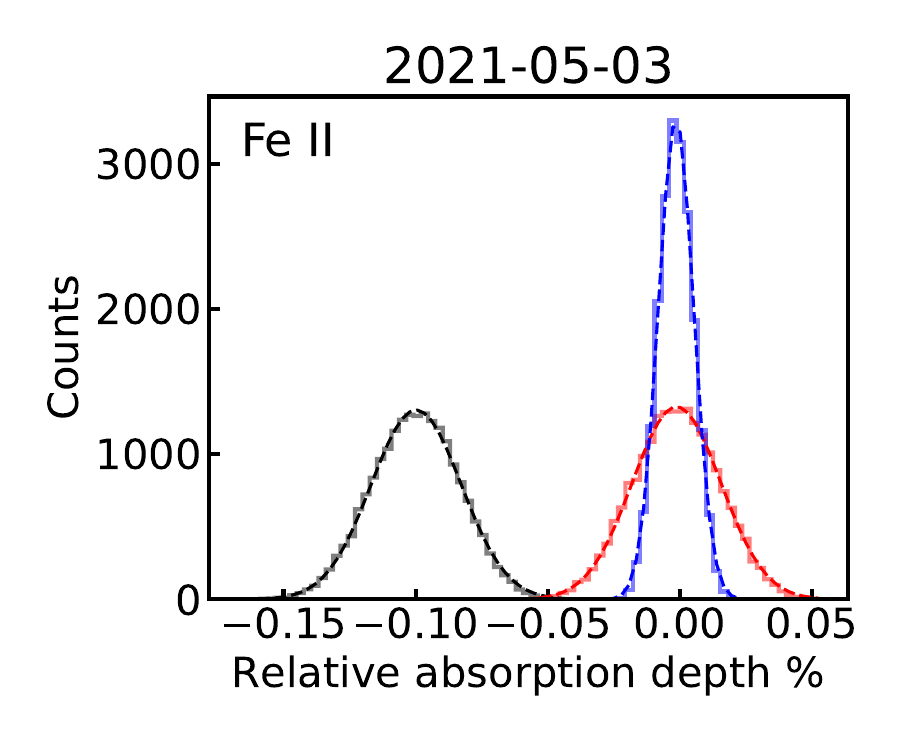}
        \includegraphics[width=6cm]{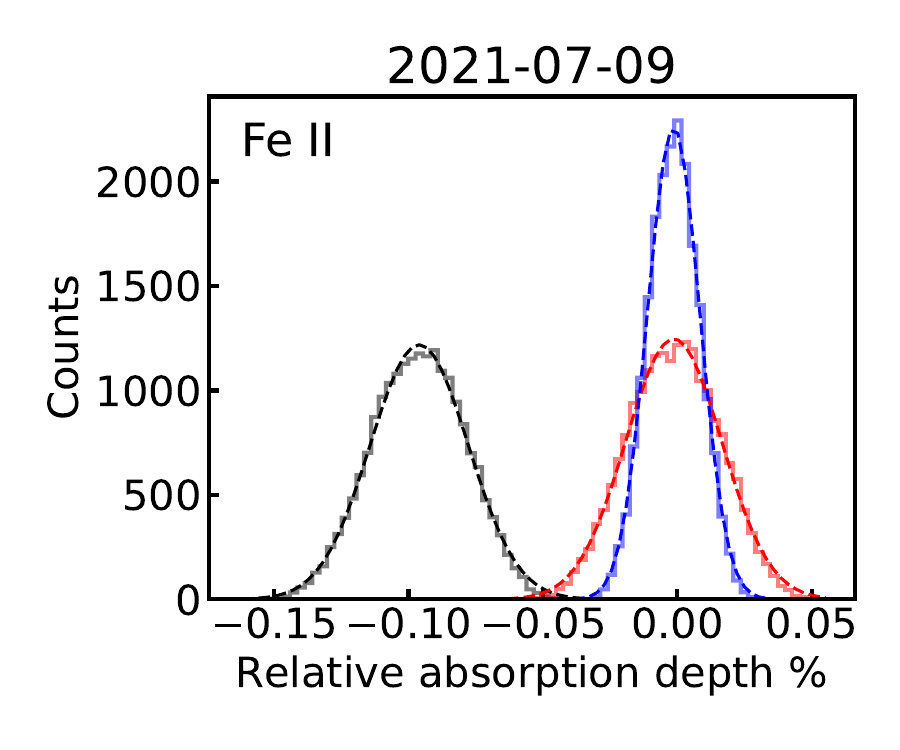}
        \caption{Distribution of bootstrapping analysis for 20,000 random selections for \MgI, \FeI, \FeII (top to bottom {panels}) for each night individually based on the cross-correlation results. Similar to Figs.\,\ref{fig:bootstrap_narrow_1}-\ref{fig:bootstrap_narrow_2}, the 'in-in' distributions (in red) and 'out-out' distributions (in blue) are supposed to be centred around 0, indicating the absence of the considered species. The 'in-out' distributions (shown in black) indicate the planetary origin of the feature. 
        }
        \label{fig:bootstrap_ccf}
    \end{figure}

\clearpage
\twocolumn
    \section{Rossiter-McLaughlin correction for cross-correlation}
    \label{app:RM}

    \begin{figure}[h]
        \centering
        \includegraphics[width=\linewidth]{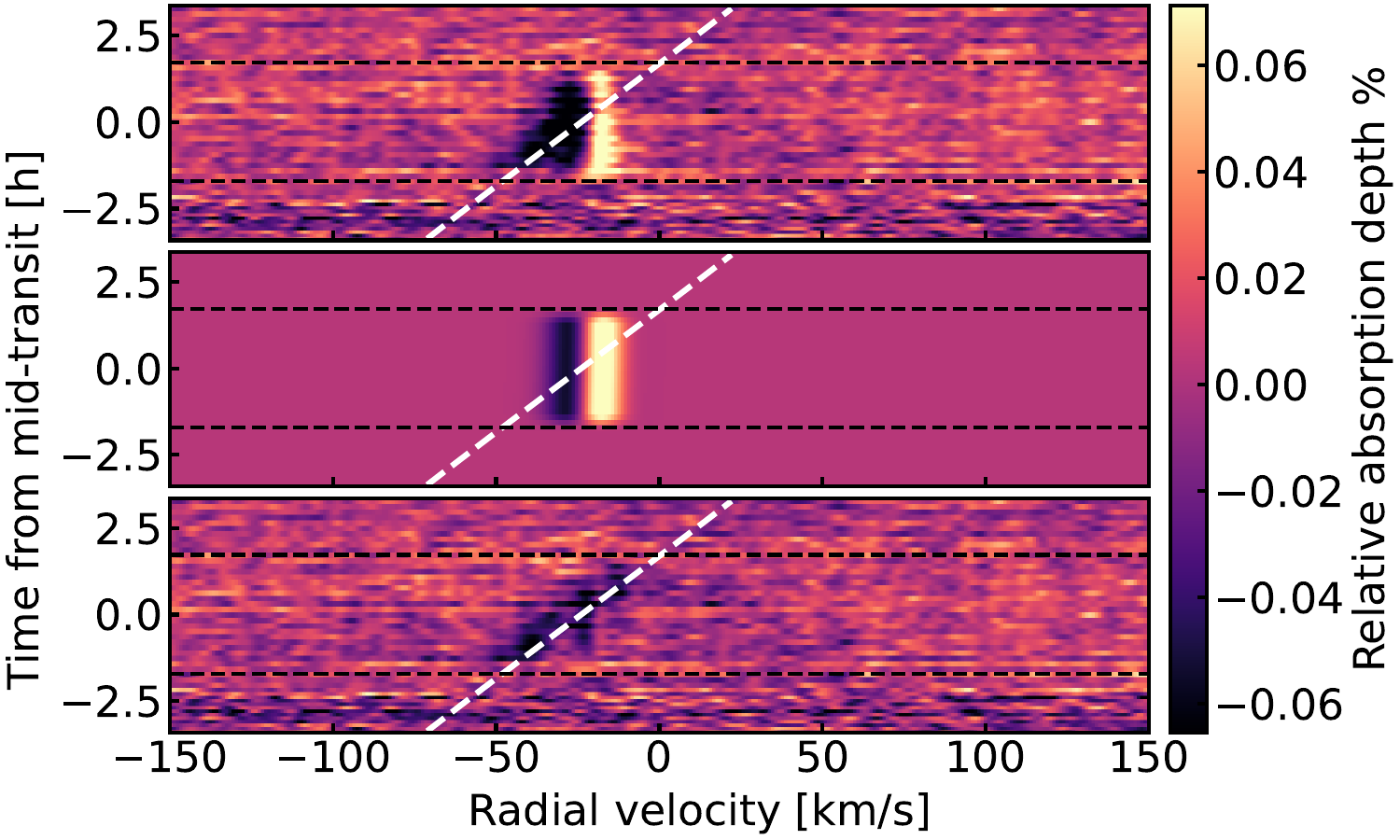}
        \caption{Rossiter-McLaughlin (RM) correction for \FeI during 2021 May 03. \textit{Top panel:} Two-dimensional cross-correlation function in the rest frame of the system. Transit contact times are indicated in black horizontal dashed lines. The expected trace of the planetary atmosphere is indicated in white. \textit{Middle panel:} Best-fit RM model. \textit{Bottom panel:} Residual after dividing the RM model. The planetary trace is now clearly visible. }
        \label{fig:RM_ccf}
    \end{figure}

    \begin{figure}[h]
        \centering
        \includegraphics[width=\linewidth]{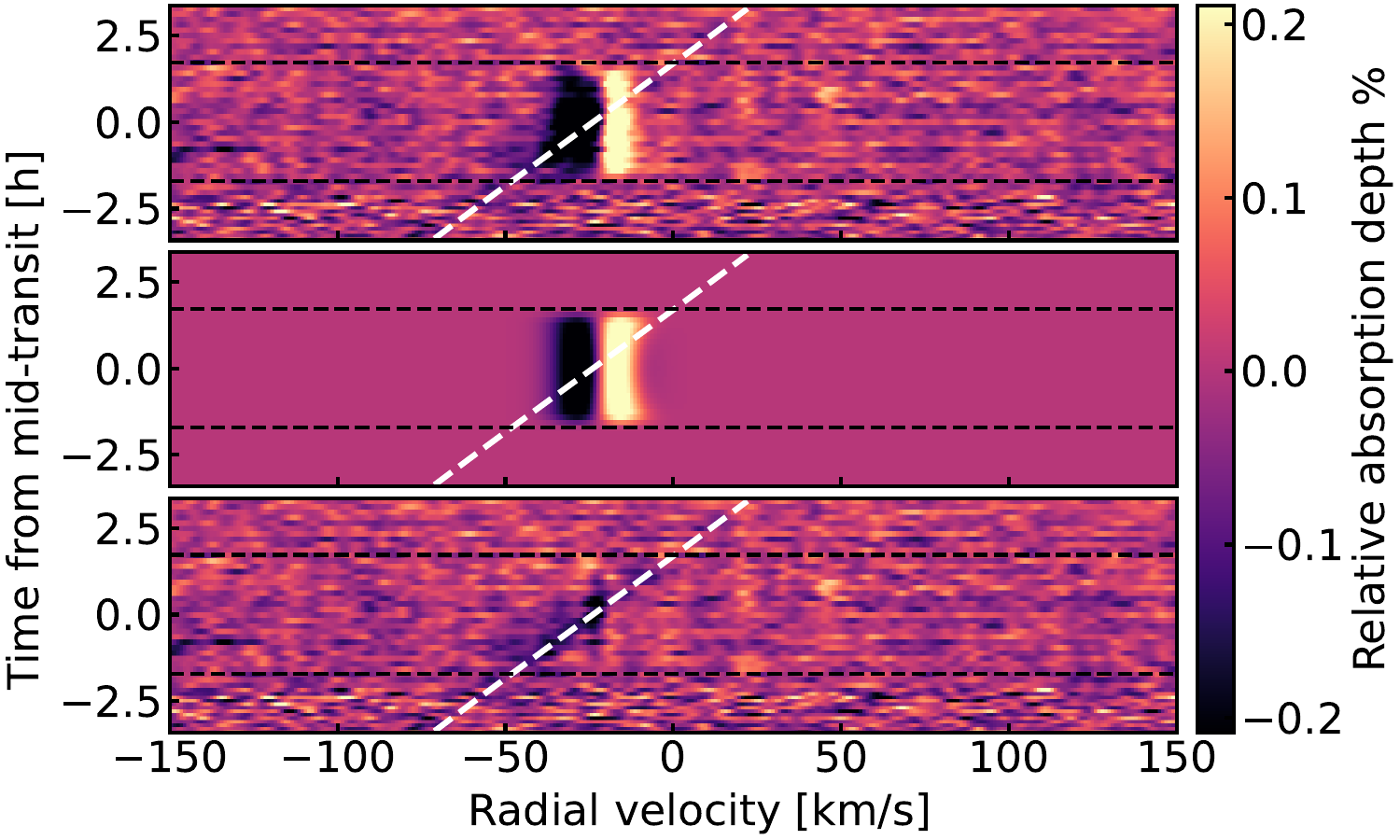}
        \caption{{Same as Fig.\,\ref{fig:RM_ccf} but for \FeII.}}
    \end{figure}

     \begin{figure}[h]
        \centering
        \includegraphics[width=\linewidth]{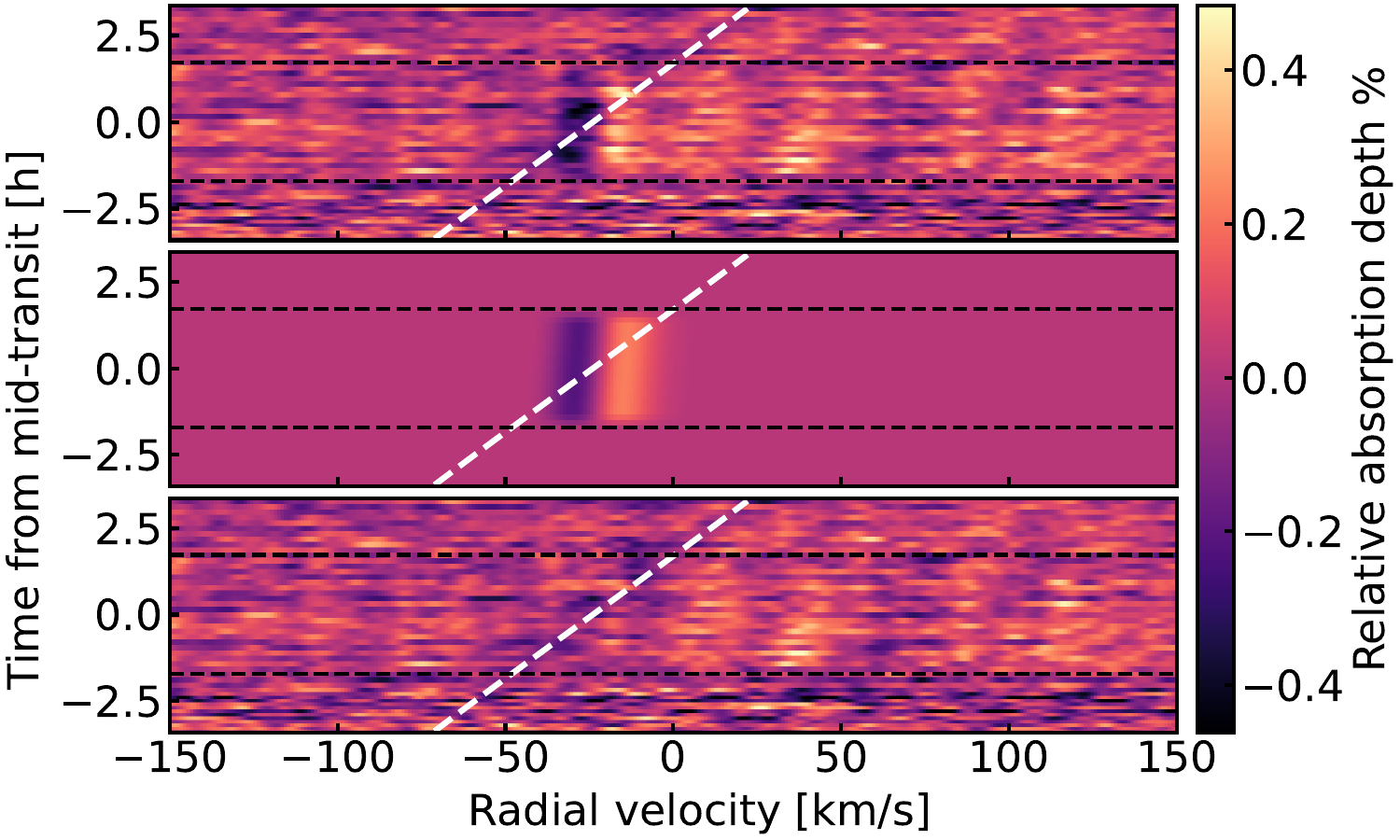}
        \caption{{Same as Fig.\,\ref{fig:RM_ccf} but for \MgI.}}
    \end{figure}
    
\end{appendix}
\end{document}